\documentclass[referee]{raa}            

\usepackage{graphicx,times}             
\usepackage{natbib}
\usepackage{amssymb,amsmath}
\bibpunct{(}{)}{;}{a}{}{,}

\usepackage[pagebackref=true]{hyperref}
\usepackage[flushleft]{threeparttable}

\begin{document}

  \title{Magnetic activity and differential rotation of HIP12653 \thanks{Based on data obtained from the ESO Science Archive Facility}}

   \volnopage{Vol.0 (20xx) No.0, 000--000}      
   \setcounter{page}{1}          

   \author{Amina Boulkaboul
      \inst{1}
   \and Lotfi Yelles Chaouche
      \inst{1}
   \and Alessandro C. Lanzafame
      \inst{2}
      \and Yassine Damerdji
      \inst{1,3}
   }

   \institute{CRAAG - Centre de Recherche en Astronomie, Astrophysique et Géophysique, Route de  l’Observatoire, Bp 63 Bouzareah, DZ-16340, Algiers, Algeria; {\it amina.boulkaboul@craag.edu.dz}\\
        \and
             Dipartimento di Fisica e Astronomia “Ettore Majorana”, Università di Catania, Via S. Sofia 64, 95123 Catania, Italy\\
        \and
             Space sciences, Technologies and Astrophysics Research (STAR) Institute, Universit\'e de Li\`ege, Quartier Agora, All\'ee du 6 Ao\^ut 19c, B\^at. B5c, B4000-Li\`ege, Belgium\\
\vs\no
   {\small Received 20xx month day; accepted 20xx month day}}

\abstract{ We present a spectroscopic and photometric study of HIP12653 to investigate its magnetic cycle and differential rotation. Using HARPS archival spectra matched with MARCS-AMBER theoretical templates, we derive the stellar parameters (Teff, logg, FeH, and vsini) of the target. The S-index, an activity indicator based on the emission of the CaII H$\&$ K lines, is fitted to determine the magnetic cycle and rotation periods. We refine the magnetic cycle period to $5799.20 \pm 0.88$ d and suggest the existence of a secondary, shorter cycle of $674.6922 \pm 0.0098$ d, making HIP12653 the youngest star known to exhibit such a short activity cycle. During the minimum activity phase, a rotation period of $4.8$ d is estimated. This is notably different from the 7-day period obtained when measurements during minimum activity are excluded, suggesting that these two periods are rotation periods at different latitudes. To explore this hypothesis, we introduce a novel light curve fitting method that incorporates multiple harmonics to model different spot configurations. Applied to synthetic light curves, the method recovers at least two rotation periods close to the true input values (within three times their uncertainties) in $92.1\%$ of cases. The inferred rotation shear shows a median deviation of $0.0011 \pm 0.0003$ and a standard deviation of $0.0177 \pm 0.0002$ from the true value.
Applying this approach to TESS photometric data from 2018 to 2023, we detect three distinct rotation periods—$4.8$ d, $5.7$ d, and $7.7$ d, (along with a signal at $3.75$ d interpreted as its first harmonic)—consistent with spots located at different latitudes. Assuming a solar-like differential rotation, we estimate an inclination of $34.0 \pm 1.8^\circ$ and a rotational shear of $\alpha = 0.38 \pm 0.01$. These results confirm the $4.8$-day period and demonstrate that differential rotation can be constrained by tracking rotation period changes across different phases of the magnetic cycle. 
}

   \authorrunning{A. Boulkaboul et al. }            
   \titlerunning{HIP12653 magnetic activity}  

   \maketitle

%
%
\section{Introduction}           
\label{sect:intro}

Late-type stars are found to exhibit magnetic cycles similar to the 11-year solar activity cycle, but their properties vary widely depending on stellar parameters such as age, rotation rate, and internal structure. While the Sun displays a relatively regular cycle, other stars can show a range of behaviors. Slowly rotating, older stars tend to have smooth and well-defined cycles, whereas younger, rapidly rotating stars often exhibit more complex, multi-periodic, or even chaotic variability \citep{baliunas1995chromospheric,olah2009multiple,olah2016magnetic}. Solar-like cycles are not ubiquitous: only about $60\%$ of solar-type stars display regular cyclic activity, while $25\%$ show aperiodic variability and $15\%$ appear magnetically flat \citep{charbonneau2012solar}.

Stellar magnetic activity in solar-like stars is a fascinating yet challenging phenomenon. There are still many unsettled questions about the underlying mechanisms involved in producing, storing, transporting, and the emergence of the magnetic field (see reviews by \cite{charbonneau2020dynamo} and \cite{charbonneau2023evolution}). The same magnetic processes are responsible for the formation and surface emergence of starspots, which are localized manifestations of strong magnetic fields inhibiting convection. These spots appear at the photosphere where magnetic flux tubes, generated and amplified in the stellar interior, buoyantly rise and emerge.

It appears that the stellar rotation plays an important role in the processes leading to the variability of the stellar magnetic field (e.g. \cite{icsik2018forward}). 
In a more general framework, recent advances in dynamo modelling indicate that the equatorial rotation period and the amount of shear towards the poles are directly impacting the stellar magnetic field temporal variability and spatial distribution (see the detailed review by \cite{kapyla2023simulations}). Observations remain a key ingredient necessary to shed further light on the ongoing physical phenomena and help constrain competing dynamo models. In this regard, the use of extreme cases, such as stars with particularly short activity cycles, can be of great benefit to testing dynamo models \citep[e.g.][]{alvarado2018far}. This highlights the importance of determining the rotation period of this type of star. Rotation periods are determined from the light curve using different methods, such as the autocorrelation function \citep{mcquillan2013measuring} and the Least-Squares periodogram \citep{nielsen2013rotation}.

Observations indicate that the strength of differential rotation depends on the stellar spectral type, as it is strongly influenced by the depth of the convective envelope. Along the main sequence, differential rotation increases with effective temperature, reaching a maximum in mid-F-type stars which possess shallower convection zones and more efficient angular momentum transport \citep{barnes2005dependence,collier2007differential}. Latitudinal differential rotation on the Sun is directly observed by tracking the motion of sunspots. In contrast, the surfaces of other stars are unresolved, so differential rotation must be inferred using indirect photometric and spectroscopic methods.

\cite{reiners2002feasibility} and \cite{ammler2012new} applied Fourier Transform analysis to the rotation line profile of Doppler-broadened spectral lines to estimate the amount of the differential rotation. In photometric data, spots at different latitudes introduce multiple periods into the light curve. \cite{lanza2014measuring} quantified differential rotation by modelling two non-evolving spots and optimising their lifetimes and rotational periods using a Markov Chain Monte Carlo approach. \cite{walkowicz2013information} synthesised light curves from spotted star models to assess degeneracies in their interpretation, highlighting that many distinct spot configurations can produce similar photometric variations. Their analysis considered scenarios both with and without differential rotation. \cite{reinhold2013fast} inferred differential rotation by iteratively fitting a global sine function to periods identified from the prewhitened Least-Squares periodogram. In each iteration, the period corresponding to the highest peak was fitted, and its resulting light curve was subtracted, 'the process known as prewhitening' by sequentially fitting and subtracting sinusoidal components. This process continued until a global fit was achieved using all selected periods.

Another method for measuring the stellar rotation period is through the S-index. Derived from emission in the Ca II H $\&$ K Fraunhofer lines, the S-index was introduced by \cite{vaughan1978flux} as a standardised approach to analyse the core fluxes of these lines in the Sun and was later extended to Sun-like stars. It is an excellent indicator of chromospheric activity \citep{hall2008stellar} and is widely used to estimate both magnetic activity cycles and stellar rotation periods.

\cite{carrington1858distribution} first identified the latitudinal variation of solar active regions during the magnetic cycle. Later, Sp$\ddot{o}$rer observed that at the start of the magnetic cycle, sunspots emerge at high latitudes between $30^{\circ}$ and $40^{\circ}$. As the cycle progresses, these spots drift toward lower latitudes, reaching around $20^{^\circ}$ at the activity maximum, and eventually migrate closer to the equator by the cycle's end \citep{hathaway2007solar}. Because the latitudinal distribution of spots changes with the magnetic cycle phase, differential rotation can be inferred from these patterns.

This paper investigates the magnetic activity and differential rotation of HIP12653. As the S-index traces the magnetic cycle, sufficient measurements taken during minimum activity can be used to estimate differential rotation. Additionally, a novel method for determining differential rotation from photometric data is introduced, specifically designed to reduce the impact of harmonic detections.

The paper is organised as follows: Section \ref{targ} provides an overview of the star under study. Section \ref{method} describes the methodology for estimating the S-index and introduces a new approach for fitting photometric data. The Monte Carlo simulations used to test this new method are described in Section \ref{mc_sim}. Section \ref{results} presents the results of the periodicity analysis applied to the S-index and TESS data. Finally, Section \ref{conc} summarises the key conclusions drawn from this study.

\section{The target star}
\label{targ}

\begin{figure}
\begin{center}
\includegraphics[width=0.8\linewidth]{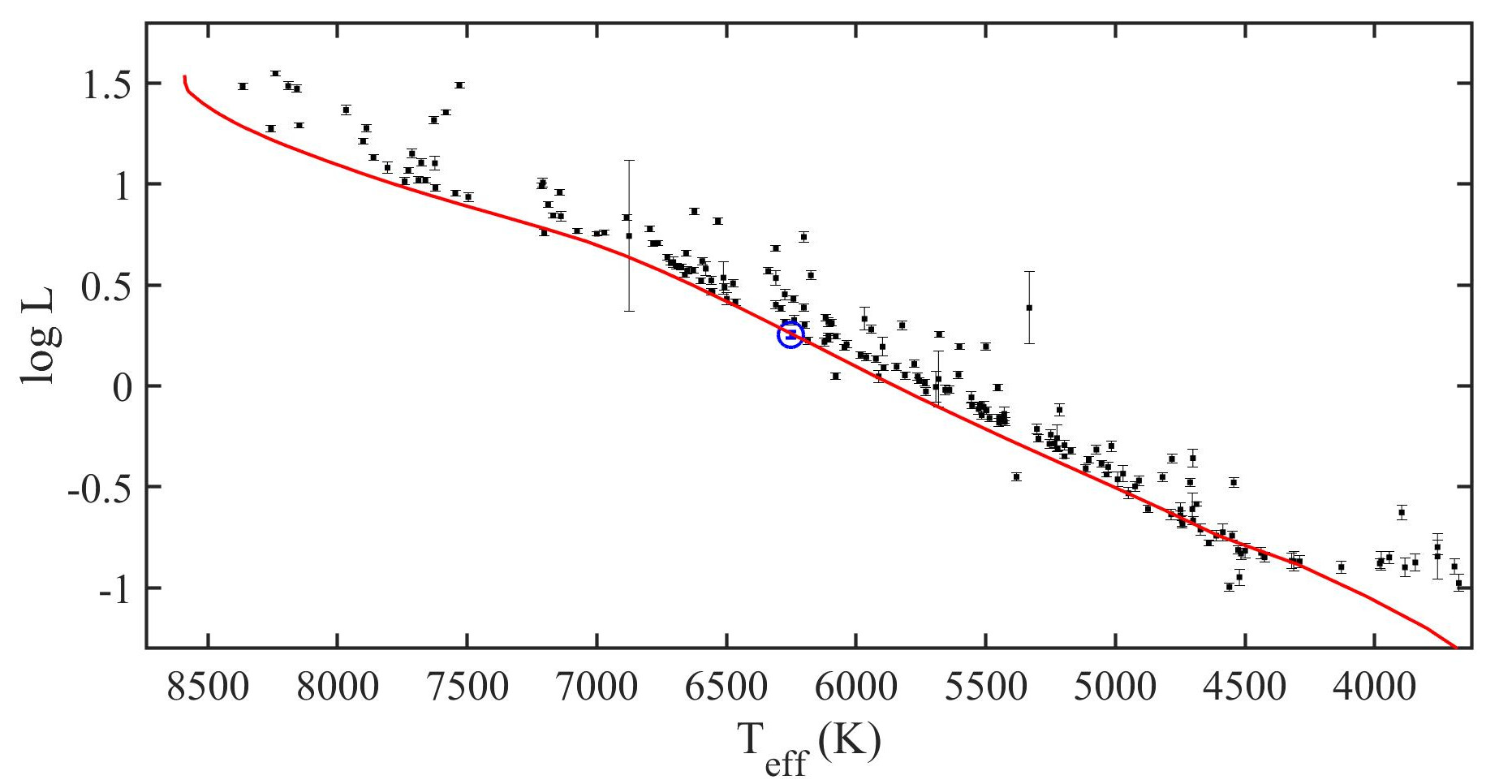}
\end{center}
\caption{HR diagram of the Hyades cluster. Black points are data from de \protect\cite{2001A&A...367..111D}, the red line is a 625 Myr isochrone with [Fe/H]=0.14 from CMD 3.8, and the blue dot marks HIP12653.}
\label{fig:isochHyad}
\end{figure}

HIP12653; $\iota$ Horologii ($\iota$ Hor) is a late type, F8V star, with a visual magnitude of $5.40$, located in the Southern hemisphere [$\alpha$ (J2000): $02h 42m 33.47s$, $\delta$ (J2000): $-50^{\circ} 48$' $01.06$'']\footnote{SIMBAD}. Studies by \cite{montes2001late, metcalfe2010discovery} suggested it to be a member of the Hyades cluster, which corresponds to an estimated age of approximately 625 Myr \citep{lebreton2001helium}. Figure\ref{fig:isochHyad} shows the HR diagram of the Hyades cluster. The black points represent observational data from \cite{2001A&A...367..111D}, the red curve corresponds to a 625 Myr isochrone with [Fe/H]=0.14 generated using CMD 3.8\footnote{\url{https://stev.oapd.inaf.it/cmd_3.8/help.html}} \citep{bressan2012parsec}, and the blue dot marks the position of our target star.
This would make it the youngest star known to exhibit a magnetic activity cycle. This star is known to harbour a $6.2\pm0.5$ Jupiter-mass planet with a period of $332\pm6$ d \citep{arenou2023gaia}. Both chromospheric and coronal (X-ray) activity cycles have been detected in HIP12653 \citep{sanz2013iotahorologi}. Previous studies have estimated different values for the star's magnetic cycle period, $P_{mag}$: \cite{metcalfe2010discovery} reported $P_{mag} = 1.6$ years, which remains the shortest coronal activity cycle known to date. In contrast, \cite{flores2016possible} found $P_{mag} = 4.57$ years. Additionally, \cite{alvarado2018far} detected two periods of 1.9 years and 1.4 years, and also estimated a rotation period of 7.7 days. 

This star is part of the radial velocity (RV) standard stars catalogue \citep{soubiran2018gaia}, used for wavelength calibration of the RV spectrometer (RVS) onboard the Gaia satellite.

This star was selected as the focus of this study due to its unique combination of properties: it is the youngest known solar-type star with a confirmed short magnetic cycle, and it exhibits one of the shortest coronal activity cycles observed to date. These features make it a valuable laboratory for probing the nature of rapid magnetic cycles in young stars. Its status as a calibration RV standard ensures long-term, high-quality, and stable spectroscopic observations, making it particularly well-suited for detailed activity analysis. Indeed, the availability of both long-term HARPS spectroscopy and TESS photometry enables a comprehensive analysis of activity signals on both magnetic and rotational timescales.

\section{METHODS}
\label{method}
This section presents the estimation of the S-index from spectroscopic data, the identification and search for periodicities in the S-index time series, and the analysis of photometric light curves to investigate stellar rotation and differential rotation.
\subsection{S-index estimate}
To estimate the S-index, we used archived spectra publicly available through the European Southern Observatory (ESO) archive\footnote{\url{http://archive.eso.org/wdb/wdb/adp/phase3_spectral/form}}. These spectra were obtained using the HARPS spectrograph, which offers a high resolving power of $R$ = $\lambda/\Delta\lambda$ = 115000. Our dataset comprises a total of 2137 spectra, spanning a timespan of 6558 days.

To accurately locate the spectral lines required for calculating the S-index, we first estimated the RVs by matching the observed spectra with synthetic spectra. Atmosphere models from MARCS \citep{gustafsson2008grid}- and AMBER \citep{de2012ambre} synthetic spectra, publicly available in the Pollux database\footnote{\url{http://pollux.oreme.org}}, were employed for this purpose. These models cover both plane-parallel and spherical geometries, with surface gravity $\log g$ ranging from $0$ to $5.5$, and a microturbulence velocity (either 1 or 2 $\rm{km \, s^{-1}}$), suitable for cool stars ranging from $3500$ to $8000$ K. The matching process relies on minimising the quadratic sum, $\chi^2$, of the differences between the observed and theoretical spectra. The methodology is detailed in \cite{boulkaboul2022analysis}, and we summarise it in what follows. We conducted iterations over MARCS-AMBRE plane-parallel and spherical synthetic spectra, varying initial atmospheric parameters (APs) from [$p - \Delta \mathrm{p}$] to [$p + \Delta \mathrm{p}$], where $p$ represents one of the APs: effective temperature $T_{\mathrm{eff}}$, surface gravity $\log\,g$, or metallicity [Fe$\slash$H]. These starting parameters were derived from \cite{soubiran2016pastel}. The step sizes for the parameters are as follows: $\Delta T_{\mathrm{eff}} = 250$K, $\Delta \log\,g = 0.5$ dex, and $\Delta$[Fe$\slash$H] = 0.25 dex. If any of these parameters reach the boundary of their interval, denoted as $p_0 = [p \pm \Delta \mathrm{p}]$, the second iteration template parameters are chosen within the extended intervals $[p, p + 2 \times \Delta \mathrm{p}]$ or $[p - 2 \times \Delta \mathrm{p}, p]$, accordingly.
Convergence in the iterative process is achieved when the parameter of interest lies at the centre of the search box. This allowed us to estimate the star's RVs and to identify the best-matching template for the target with atmospheric parameters and projected rotation velocity (vsini). The resulting parameters are compared to those reported by \cite{soubiran2016pastel}, as shown in Table. \ref{tab:APS}. We note that a precise inference of the APs is beyond the scope of this work; for the purposes of our analysis, an estimate within the spacing of the grid is sufficient.

\begin{figure}
\begin{center}
\includegraphics[width=0.8\linewidth]{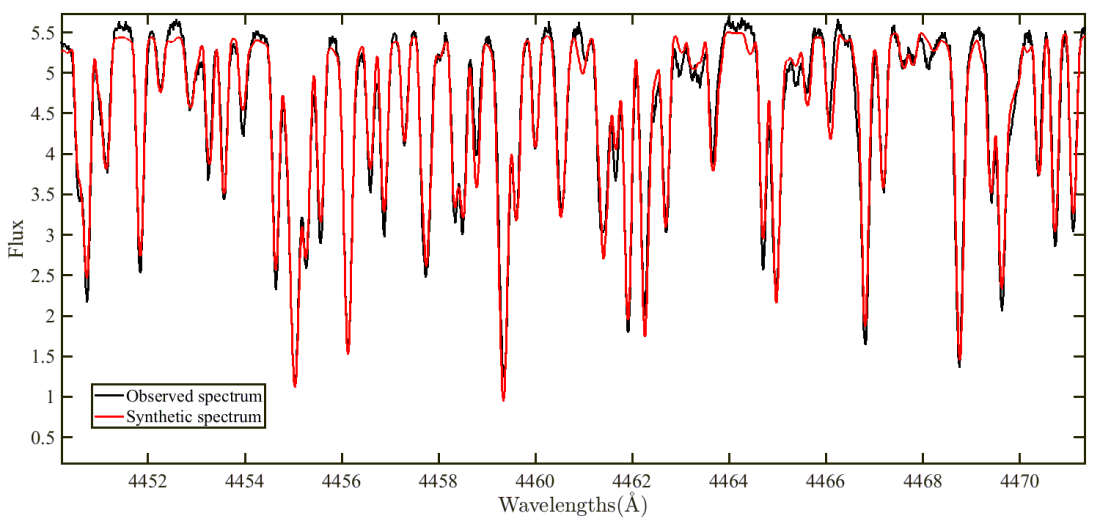}
\end{center}
\caption{HARPS spectrum of HIP12653 for observation at BJD = 2454059.5 (black) in the spectral range $4450-4471 $\AA, with the best-fit MARCS-AMBRE synthetic spectrum overplotted in red.}
\label{fig:spectrum}
\end{figure}

\begin{table}
 \begin{center}
   \caption{\small Stellar parameters derived from the best-fit synthetic spectrum—$T_{\mathrm{eff}}$, $\log g$, [Fe/H], [$\alpha$/Fe], and $v \sin i$—along with corresponding values from \citet{soubiran2016pastel} for comparison.}
    \label{tab:APS}
    \begin{tabular}{rrrrrrr}
    \hline
    &{$T_{\mathrm{eff}} [K]$} & {$\log\,g$} & {[Fe$\slash$H]} & {[$\alpha\slash$Fe]} & {vsini [$\rm{km\, s^{-1}}$]} \\
\noalign{\smallskip}
\hline
\noalign{\smallskip}
This work & $6250$ & $4.50$ & $0.0$ & $0.0$ & $6.84 \pm 0.36$ \\
\protect\cite{soubiran2016pastel} & $6239$ & $4.55$ & $0.12$& -- & -- \\
\hline
\end{tabular}
\end{center}
\end{table}
An example is provided in Fig. \ref{fig:spectrum}, which shows a fitted HARPS spectrum (black) with overplotted best-fit synthetic spectrum (red) in the spectral range $4450-4471 $\AA. The fit is generally good, with the exception of a few weaker lines. These discrepancies may be attributed to template mismatches and the limitations of the line list used to generate the synthetic spectrum. The parameters for the shown HARPS spectrum are summarized in Table \ref{tab:bestFit} These include the signal-to-noise ratio (SNR), the measured RV, and the reduced chi-squared ($\chi^2$) of the best-fit model.
\begin{table}
 \begin{center}
   \caption{\small Parameters of the HARPS spectrum of HIP12653 (BJD = 2454059.5) in the spectral range $4450-4471 $\AA: signal-to-noise ratio (SNR), measured RV, and reduced chi-squared ($\chi^2$) of the best-fit synthetic spectrum.}
    \label{tab:bestFit}
    \begin{tabular}{rrrr}
    \hline
    {BJD} & SNR & $RV [km s^{-1}]$ & $\chi^2$  \\
\noalign{\smallskip}
\hline
\noalign{\smallskip}
$2454059.5275$ & $145.2$ & $17.039 \pm 0.021$ & $108.7$ \\
\hline
\end{tabular}
\end{center}
\end{table}

The measured RVs are used to shift spectra in order to measure the total flux inside the chromospheric emission lines CaII H $\&$ K; which gives a measurement of the S-index. We estimate the S-index using the method described by \citet{lovis2011harps}. This involves calculating the ratio of the sum of fluxes in two intervals of width 1.09 \AA\  centered on the $K$ ($3933.66$ \AA) and $H$ ($3968.47$ \AA) lines, to the sum of the fluxes in two reference windows of 20 \AA\ width, centered on the adjacent continuum passbands $V$ ($3900.0$ \AA) and $R$ ($4000.0$ \AA). The S-index is given by:
\begin{equation}
S = \alpha \frac{H+K}{R+V}
\end{equation}
$\alpha$ is a calibration constant set to 2.4 to match the Mount Wilson scale, as specified by \citet{duncan1991ii}. This value is widely used in HARPS data \citep[e.g.,][]{lovis2011harps}.

\subsection{S-index time-series fitting}
A period search is conducted on the S-index dataset using the \citet*{heck1985period} periodogram, as updated by \cite{zechmeister2009generalised}  in the form of the Generalised Least-Squares (GLS) periodogram. The periodogram settings are as follows:

The lowest frequency is set to the inverse of the time span $T$ multiplied by the factor $1.5$ while the highest frequency corresponds to the lowest period of 2 days in order to avoid the one day period aliases. The frequency step is taken as $\frac{1}{100 \,T}$.  

The empirical cumulative probability distribution function (ECDF), which allows us to set a false alarm probability (FAP) to the periodogram, is constructed from the highest peaks of $10^6$ realisations of white noise periodograms with the same time sampling. Peaks in the frequency periodograms that are over a $95\%$ probability threshold were considered significant. 
In the case where the period exceeds the observation time span, a polynomial trend is fitted and subtracted. The periods are extracted iteratively then all the solutions are refined simultaneously. 

Stellar activity cycles are not perfectly periodic \citep{olah2009multiple}. In fact, the length of individual solar cycles fluctuates between 9 and 13 years, with a standard deviation of approximately 14 months \citep{hathaway2007solar}. Additionally, the solar activity rises faster than it falls, which makes the cycle shape deviates from a sine function. 

Therefore, to model the temporal evolution of the S-index, S(t), we adopt a functional form inspired by the Keplerian RV equation, $V = V_0 + K[e cos\omega + cos(\omega + \nu)]$. We use this model to fit the S-index time series, $S = S_0 + A[e \cos\omega + \cos(\omega + \nu)]$, adapting the orbital parameters to describe features of the magnetic activity cycle rather than orbital motion. In this formulation, the period in the model represents the dominant periodicity in the S-index data. This could correspond to either: The magnetic activity cycle period if we are analysing long-term variations in the S-index, or the rotation period of the star if the variations are modulated by starspots or plages rotating in and out of view. The eccentricity, $e$, represents the asymmetry of the modulation, describing how the rise and fall of the magnetic activity deviate from a purely sinusoidal variation, indicating whether one phase (rising or decaying) is more prolonged than the other, with sharper rises or falls depending on the value of $\omega$. The parameter $A$ represents the amplitude of the S-index variation, i.e., the strength of the activity modulation, rather than an RV semi-amplitude. Phase offset, $T_0$, defines the reference time for the phase of the activity cycle, the time of minimum or maximum magnetic activity, setting the "starting point" of the cycle in time, similar to periastron passage in an orbit. This modeling allows us to account for the often asymmetric and quasi-periodic nature of stellar magnetic activity cycles, as observed in the Sun.

To estimate these parameters, we use the significant frequency peaks as input to a Zechmeister \& Kürster Keplerian periodogram. For each significant frequency $f_i$, we step over a grid of asymmetry factors $e$ from 0 to 0.95 and phase offsets $T_0$ in the range $[0, 0.95 f_i]$, and fit the S-index time series with the model: $S(t) = a (\cos \nu + e) + b \sin \nu$ where the true anomaly $\nu$ is computed from $f_i$, $e$, and $T_0$. The coefficients $a$ and $b$ are linear parameters, with $a = A \cos \omega \quad \text{and} \quad b = -A \sin \omega$. Finally, all parameters are refined using the Levenberg--Marquardt algorithm. The uncertainties in these parameters are obtained from the variance–covariance matrix.

\subsection{Photometric data fitting}
\subsubsection{Description}
The presence of active regions on the photosphere leads to photometric variability in the star's light curve, modulated by the star's rotation period. A key driver of the solar dynamo is differential rotation. During the magnetic cycle, starspots emerge at varying latitudes, each with distinct rotation periods due to this differential rotation. The configuration of spots across the stellar surface generates a composite light curve signal that includes contributions from rotation periods at different latitudes and their harmonics. Consequently, distinguishing the harmonics from the primary rotation periods is essential for accurate analysis.

This method aims to determine differential rotation, which should not be confused with harmonics of the main period. 

Our fitting model starts by applying the GLS periodogram to the TESS light curve (see also sect. \ref{tess}), using a frequency lower limit given by the inverse product of the observation time span and an oversampling factor, which we set to 1.5. The period corresponding to the highest peak in the periodogram represents the dominant signal in the data. Other periods may arise from harmonics of the main period, contributions from other spots (as stars are typically not covered by a single spot), or sampling effects. Simultaneously fitting all periods from the periodogram may yield a model that closely matches the data but does not necessarily correspond to distinct physical spots.

To address this, we fit the primary light curve using a sine function with the dominant period (highest peak) and include different numbers of harmonics $n$: $f = c + \sum_i^n{a_i\cos(2\pi i t/P) + b_i\sin(2\pi i t/P)}$. The fitting parameters, $a_i, b_i, c$, are determined through $\chi^2$ minimization. The optimal number of harmonics, $n$, introduced to account for potential $180^{\circ}$-longitude separation configurations, is determined using a Fisher test. This approach ensures that harmonics of the dominant period are captured in the primary fit and do not appear as artefacts in the residuals. 
The fitted signal is then subtracted from the light curve, and GLS periodogram is reapplied to the residuals to extract the next highest peak period. This newly identified period, along with the previously determined one, is then simultaneously fitted using sine functions with different numbers of harmonics. If the inclusion of the new period does not significantly improve the fit based on a Fisher test, the process is terminated, retaining only the initial dominant period. This iterative process continues, with each newly identified period and all previously determined periods being refitted simultaneously. The procedure is halted when the root mean square error (RMSE) relative to the signal amplitude no longer decreases. The number of iterations should be as large as possible, but in the case of simulated data, it should be limited to the number of spots in the model.

The highest peak in the periodogram may correspond to the first harmonic of the true rotation period rather than the rotation period itself, due to the presence of two active longitudes separated by approximately $180^{\circ}$. To identify the correct period in observational data, we compare the highest peak period in the initial periodogram ($P_1$) with half the period of the second-highest peak ($P_2$). If $P_1$ and $P_2$ differ by less than $5\%$, and the highest peak in the residuals periodogram is distinct from $P_2$, $P_2$ is considered as the primary period.

Once all the periods and their best number of harmonics are selected, the periods are optimised with the Levenberg Marquardt algorithm to estimate their uncertainties. 

\subsubsection{Degeneracies in light curve}
Our approach does not aim to reconstruct detailed spot topography or derive latitude-dependent rotation rates. Instead, it focuses on detecting multiple periodic signals in the photometric time series, each interpreted as the rotation period of a distinct active region (AR). We explicitly define differential rotation in this context as the difference between the maximum and minimum rotation periods detected from the light curve. These periods are attributed to ARs, regardless of their specific latitudes or stellar inclination. Thus, our results reflect the differential rotation of the activity belt, rather than a full surface map. Following methodologies similar to \cite{reinhold2013fast} and \cite{balona2016differential}, we measure differential rotation as the spread between distinct rotation period signals detected in the photometric time‑series. These approaches do not require spot latitude determination or inversion modeling; they yield robust lower limits on surface shear using purely timing-based diagnostics.

The large number of free parameters in light curve inversion (e.g., inclination, number of ARs, their locations, and sizes) introduces degeneracies, whereby multiple spot configurations can produce the same photometric variability \citep{walkowicz2013information}. Since an accurate determination of differential rotation would require knowledge of spot latitudes, which is only possible via Doppler Imaging \citep{vogt1983doppler}, we provide only a lower limit on the surface shear. Rather than inferring spot location or contrasts, we extract a series of independent periodicities through iterative sinusoidal fitting and periodogram analysis, combined with statistical tests to avoid overfitting and distinguish physical signals from harmonics or noise. Although the true spot latitudes are not retrieved, we argue that the range of recovered rotation periods still carries meaningful information about latitudinal shear within the activity belt. The inclination angle of the stellar rotation axis affects the detectability of spot-induced modulations, with low-inclination (i.e., pole-on) views reducing sensitivity to low-latitude spots. As a result, the spread in observed rotation periods may underestimate the full latitudinal shear. Nevertheless, the detected periods correspond to the true rotation periods of visible AR, and the inferred differential rotation reflects a lower limit on the shear within the activity belt.

Most stars do not exhibit spot coverage from equator to pole; instead, magnetic activity tends to concentrate within specific latitude belts, as in the Sun, where ARs are confined to around $\pm30^{\circ}$ during most of the cycle. By measuring the range of rotation periods of these ARs, we obtain an estimate of the differential rotation within the activity belt. For stars with long-term photometric monitoring, tracking how this period spread evolves over time may offer further insight into the latitudinal migration of activity, even in the absence of direct latitude measurements.

\section{Monte Carlo simulation}
\label{mc_sim}
\subsection{Description}
To simulate the effect of spots on the light curve, we use the SOAP software \citep{boisse2012soap}, which computes the flux of a spotted star. We generate 10000 light curves, each representing a unique realization of a star with ARs. The stellar parameters varied for each iteration include the inclination, $i$, the amount of differential rotation shear $\alpha$, the rotation period, the number of ARs, their latitudes, starting longitudes and their sizes. The amount of differential rotation shear is given by :
\begin{equation}
    \alpha = \frac{P_{pole} - P_{eq}}{P_{pole}}
\end{equation}
where $P_{pole}$ and $P_{eq}$ are the rotation periods at the poles and the equator, respectively. For all iterations, the equatorial period was fixed at 4 days, consistent with measurements for HIP12653 during its activity minimum phase (see Sect. \ref{sect:RotSindex}).

For each iteration, the parameters were randomly sampled from uniform distributions. The number of ARs was drawn from the interval $[2, 6]$. 

The $\alpha$ value, drawn between $[0.1, 0.7]$ range, encompasses values measured or inferred for solar-type and hotter stars. Observational studies of F-type stars (e.g., \citealt{balona2016differential}) suggest they can exhibit relatively strong differential rotation, with $\alpha$ reaching up to $0.5-0.6$ in some cases \citep{ammler2012new}, especially for stars with shallow convective zones. This choice allows us to explore both solar-like and stronger-than-solar shears.

For each AR, the latitude base was randomly drawn between the equator and the poles [0, $90^{\circ}$], while their size was randomly chosen between 0.05 and 0.15 in units of stellar radius. This corresponds to filling factors (spot area divided by visible disk area) ranging from $f = 0.25\%$ to $f = 2.25\%$ for each AR, this amounts to a total filling factor ranging from $0.5\%$ to $13.5\%$. These values are consistent with the range of spot coverage observed in solar-like and F-type stars. On the Sun, the maximum sunspot filling factor during activity maximum rarely exceeds $1\%$ \citep{solanki2003sunspots} and \cite{araujo2025starspot} found a spot coverage of $5\%$ for the F5-type star CoRoT-6. 

The inclination angle, $i$, was sampled uniformly between 10 and 50 degree. This range was chosen to avoid the extremes of nearly pole-on or equator-on geometries, both of which present challenges in interpreting light curve modulation. At very low inclinations, active regions produce minimal variability due to their limited apparent motion across the stellar disk. On the other hand, very high inclinations can lead to strong modulation, but also introduce complex visibility patterns and projection effects that can complicate the signal interpretation. By focusing on moderate inclinations, we ensure detectable rotational modulation while maintaining a realistic geometric configuration.

The simulation spanned fixed time durations of 30 and 60 days to evaluate the effect of different observation windows. Each AR was confined to a belt that lasted three rotation periods, accounting for its decay, and the appearance of another AR at close latitude and different longitude, which can create a harmonic of the rotation period. The rotation period at a given latitude, $P(l)$, was computed using the relation:
\begin{equation}
    P(l) = \frac{P_{eq}}{1 - \alpha\sin^2(l)}
\end{equation}

Here, $l$ is the AR's latitude, which was varied from the latitude base by a random value between $-2^{\circ}$ and $+2^{\circ}$, confining the ARs around the same latitude. The resulting flux is the sum of all individual ARs contributions generated by SOAP for the specified inclination, $\alpha$, and rotation period. This approach effectively simulates the presence of multiple ARs at varying latitudes with defined lifetimes.

\section{Results and Discussion}
\label{results}
\subsection{HIP12653 Chromospheric activity}
\label{chromo}
\begin{figure}
\begin{center}
\includegraphics[width=0.8\linewidth]{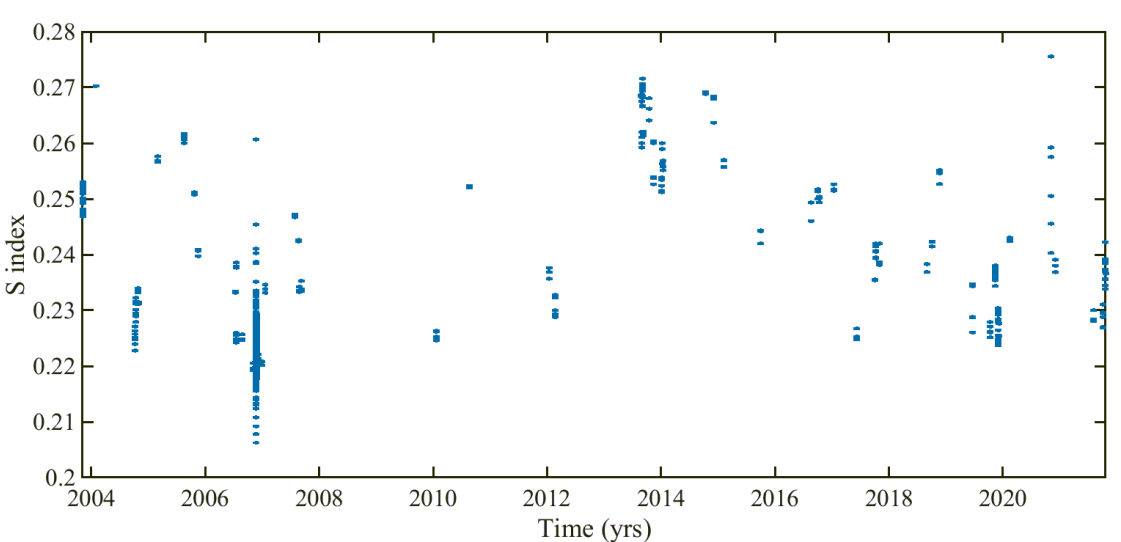}
\end{center}
\caption{Variation of the S index of HIP12653 as a function of time.}
\label{sindexData}
\end{figure}

The S-index data covers a duration of 6558 days, from 2003 to 2021. Figure \ref{sindexData} shows the variation of the S-index as a function of time, while the individual measurements are provided in Table \ref{tab:sindexMes}, revealing a clear periodic signal associated with the star's magnetic cycle, with a minimum occurring around 2006. To identify the significant periodic signals present in the S-index data, we applied the GLS periodogram. Table \ref{tab:periodsALL} lists the detected periods along with the amplitudes $A$, the factor of asymmetry $e$, and the maximum activity timing $T_0$.

\begin{table}
\centering
\begin{threeparttable}
   \caption{\small S-index measurements.}
    \vspace*{5mm}
    \label{tab:sindexMes}
    \begin{tabular}{rrrr}
    \hline
    {Date [BJD]} & {S-index} & {$\sigma_{S-index}$} \\
\noalign{\smallskip}
\hline
\noalign{\smallskip}
2452949.69072793 & 0.24731 & 0.00023\\
2452949.69157281 & 0.24936 & 0.00023\\
2452949.69241770 & 0.24792 & 0.00023\\
2452949.69350563 & 0.24797 & 0.00023\\
2452949.69481346 & 0.25115 & 0.00023\\
\multicolumn{3}{l}{...}\\
\hline
\end{tabular}
\begin{tablenotes}
      \footnotesize
      \item Note: This table is available in its entirety online at the CDS. A portion is shown here for guidance regarding its form and content.
    \end{tablenotes}
\end{threeparttable}
\end{table}

\begin{table*}
 \begin{center}
   \caption{\small Periods detected in the S-index dataset along with the signal amplitude $A$, the asymmetry factor, $e$, and the cycle timing $T_0$.}
    \vspace*{5mm}
    \label{tab:periodsALL}
    \begin{tabular}{rrrr}
    \hline
   & {$P_{mag1} $} & {$P_{mag2} $} & {$P_{rot} $} \\
\noalign{\smallskip}
\hline
 \multicolumn{4}{c}{All measurements} \\
\hline
\noalign{\smallskip}
P [d] & $5799.20 \pm 0.88$ & $674.6922 \pm 0.0098$ & $4.7699949 \pm 0.0000021 $  \\
A & $16.295 \pm 0.013$  & $11.8967 \pm 0.0055$ & $3.2945 \pm 0.0021$ \\
e & $0.62849 \pm 0.00029$ & $0.34382 \pm 0.00012$ & $0.029734 \pm 0.000019$ \\
$T_0$ [BJD] & $2456391.10 \pm 0.67$ & $2454795.494 \pm 0.057$ & $2454404.17106 \pm 0.00044$ \\
\hline
 \multicolumn{4}{c}{Minimum activity measurements excluded}\\
\hline
\noalign{\smallskip}
P [d] & $4453.82482 \pm 0.00040$ & $664.357 \pm 0.012$ & $7.6383684 \pm 0.0000089 $  \\
A & $17.0262 \pm 0.0070$  & $14.6659 \pm 0.0094$ & $2.9750 \pm 0.0062$ \\
e & $0.401637500 \pm 0.000000098$ & $0.60763267 \pm 0.00000015$ & $0$ \\
$T_0$ [BJD] & $2456479.7929 \pm 0.0015$ & $2456525.643 \pm 0.024$ & $2456400.6161 \pm 0.0025$ \\

\hline
\end{tabular}
\end{center}
\end{table*}
\subsubsection{Magnetic cycle}
Although the observational baseline spans approximately 18 years, slightly longer than the identified period of 5799 d (15.9 yrs), we still report this long-term signal, most likely associated with the star's magnetic cycle, as it is consistently recovered in our analysis. While the frequency resolution of the dataset ($1/T$) allows for detection of such long periods, we acknowledge that identifying a periodicity longer than half the observational timespan carries significant uncertainty. We have accounted for this limitation in our error estimation and emphasise that this detection requires further observational confirmation.

\begin{figure}
\begin{center}
\subfloat{\includegraphics[width=0.8\linewidth, trim={0 2.5cm 0 0},clip]{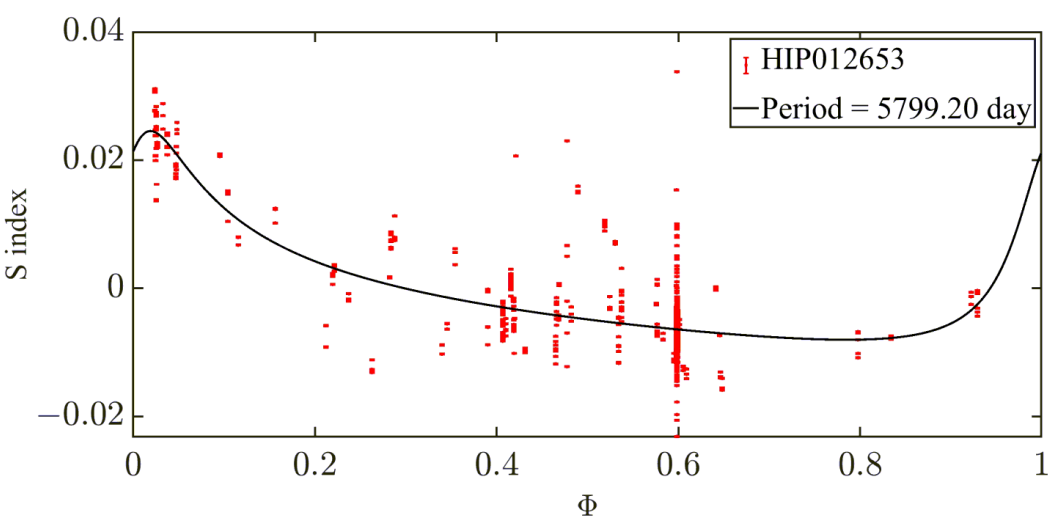}}\\
\subfloat{\includegraphics[width=0.8\linewidth]{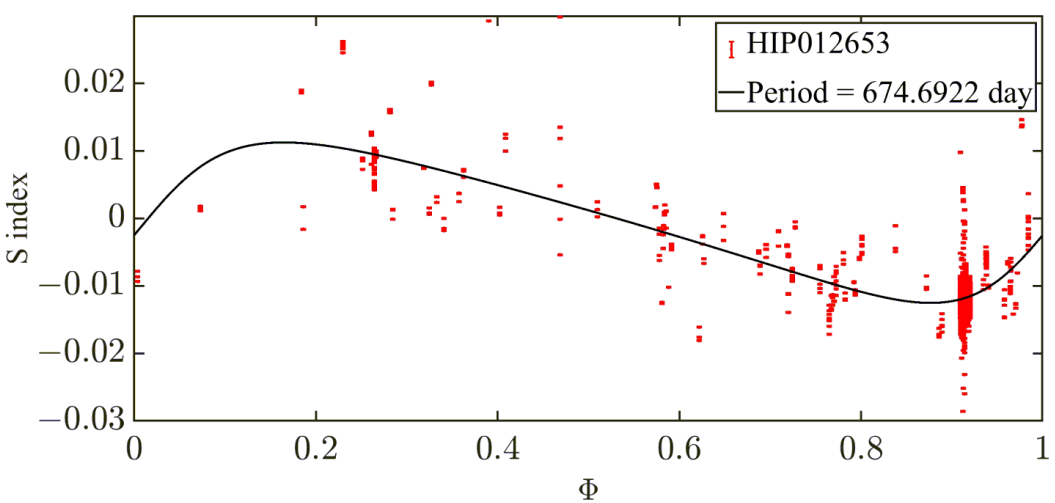}}
\end{center}
\caption{S-index data phase-folded over the periods 5799 d (top panel) and 674.7 d (bottom panel).}
\label{SindexMagP1}
\end{figure}

This value is much longer than the cycle periods previously reported in the literature: 4.57 yr estimated by \cite{flores2016possible}, 1.9 yr suggested by \cite{alvarado2018far}, and later revised to 1.5 yr by \cite{amazo2023far}. The discrepancy indicates that the magnetic cycle could be longer, particularly with the inclusion of new observations. This signal has a high asymmetry factor, $e = 0.6$ showing that the decay phase is more prolonged than the rising phase as seen in Fig. \ref{SindexMagP1} (top panel). After removing the contribution of the 5799-day period, an additional period of 674.7 days (1.84 yr) is detected (bottom panel of Fig. \ref{SindexMagP1}). This value is close to the 1.9-year cycle proposed by \cite{alvarado2018far} and $684$ d by \cite{flores2016possible}. 

If the 674.7-day period is not a harmonic of the long magnetic cycle, there are two possible explanations for its presence. First, as suggested by \cite{alvarado2018far}, the uneven spatial and temporal distribution of chromospherically active regions could lead to two signals with shorter and longer periods. These signals may arise from the non-visibility of active regions in one hemisphere at certain points in the cycle. Additionally, \cite{sanz2013iotahorologi} proposed that changes in the cycle period could be due to asymmetries in the emerging activity between the visible and partially visible hemispheres.

Second, this period could represent a secondary cycle. Numerous studies have demonstrated that stars with moderate activity levels often exhibit multiple activity cycles (e.g., \cite{olah2009multiple}, \cite{olah2016magnetic}). \cite{sanz2013iotahorologi} previously suggested that HIP12653 might exhibit multiple cycles, estimating a long period of approximately 6 years and modulating a shorter one. \cite{bohm2007chromospheric} investigated the relationship between magnetic cycle periods and stellar rotation periods, concluding that the magnetic cycle period increases with rotation period, following two distinct sequences: active and inactive branches. They proposed that each sequence, characterised by a nearly constant ratio of rotation periods to activity cycle length, corresponds to a different type of stellar dynamo. They also found that stars which belong to the active branch show two activity cycles, where most of the secondary cycles fall in the inactive branch.

\cite{alvarado2018far} derived cycle-to-rotation ratios of 67 for their short cycle and 93 for their long cycle, suggesting that only the short cycle aligns with the inactive branch. Considering our derived magnetic cycle periods and the rotation period of 7.70 d from \cite{alvarado2018far}, we calculate ratios of 753 for the long period and 87 for the short period. The ratio for the long period exceeds the 500 rotations per cycle typical of the active branch reported in \cite{bohm2007chromospheric}. This is expected, as their sample did not include stars with such long magnetic cycles. In contrast, the short-period ratio is close to the 90 rotations per cycle associated with the inactive branch, suggesting that the short period may correspond to a secondary cycle.

\subsubsection{Rotation period}
\label{sect:RotSindex}
The residuals of the S-index data, after removing the contribution of the two magnetic cycle periods, reveal a period of 4.77 d. This differs from previous values reported in the literature. \citep{saar1997rotation} determined a rotation period of 7.9 d based on the chromospheric emission level (log$R_{HK}$) and the empirical period–activity relation from \cite{noyes1984rotation}. Using S-index measurements from early 2008 to mid-2010, \cite{metcalfe2010discovery} found a period of 8.5 d, consistent with the mean rotation period of solar-type stars in the Hyades \citep{radick199512}. Meanwhile, \cite{alvarado2018far} estimated a period of 7.7 d from the mean longitudinal magnetic field. Using various methods—GLS, Gaussian-process regression, autocorrelation function, and wavelet power spectra—\cite{amazo2023far} analysed TESS photometric data (sectors S2 and S3) and found a rotation period ranging from 6.6 to 7.9 days.

The S-index data include 1820 measurements collected between November 20 and 28, 2006, coinciding with the activity minimum. Applying the GLS periodogram to this subset, we obtained the period of $4.75574 \pm 0.00088$ d. Since most of these measurements are concentrated within this week, this period is expected to dominate the period search across the entire dataset.

This period was overlooked in the analysis by \cite{alvarado2018far} because they used only one measurement per day. To verify that our results are consistent with theirs, we applied the same approach to the 2006 activity minimum, obtaining a period of $7.450033 \pm 0.000013$ d from the entire dataset.

To further confirm this, we excluded the minimum activity measurements from the full dataset analysis and derived the two long periods $4453.82482 \pm 0.00040$ d, and $664.357 \pm 0.012$ d (listed in Table \ref{tab:periodsALL}). We notice that the long magnetic cycle period changes since we have less measurements, and the time at which the maximum of activity happens is therefore different by a few months. We were able to recover a rotation period of $7.6383684 \pm 0.0000089$ d and its first harmonic $3.7536551 \pm 0.0000020$ d, which is consistent with values reported in the literature. This result suggests that the rotation period at the activity minimum differs from the period observed at other phases of the magnetic cycle due to differential rotation.

The Sun Maunder’s ‘Butterﬂy Diagram” shows a change of the latitudinal position of sunspots with time. At the start of the solar activity cycle, a few spots appear at high latitudes, migrate toward mid-latitudes as the activity reaches its maximum, and eventually approach the equator by the cycle's end. This progression demonstrates that spots at different phases of the magnetic cycle trace different latitudes. Assuming HIP12653 exhibits solar-like differential rotation, a short period observed at minimum activity suggests that, between November 20 and 28, 2006, the spots were located near the equator, marking the end of the first cycle.

We compare in Table \ref{tab:resultLit} our estimated rotation and magnetic cycle periods to those reported in literature. We note that in the periodogram of the S-index shown in the top panel of Figure 6 in \cite{alvarado2018far}, there is a peak near half the value of their strongest signal at 7.7 days. This secondary peak was not discussed in their work, as they focused solely on the dominant period.

\begin{table}
 \begin{center}
   \caption{\small Comparison of periods detected in the S-index dataset from this study and those of literature.}
    \vspace*{5mm}
    \label{tab:resultLit}
    \begin{tabular}{cccc}
    \hline
    {This work} & {\protect\cite{metcalfe2010discovery}}& \protect\cite{flores2016possible}&\protect\cite{alvarado2018far}  \\
\noalign{\smallskip}
\hline
\noalign{\smallskip}
 $4.7699949 \pm 0.0000021$ & -- & --& -- \\
 $7.6383684 \pm 0.0000089$ & $8.5 \pm 0.1$& -- & $7.70^{+0.18}_{-0.67}$ \\
 $3.7536551 \pm 0.0000020$ & --& -- & --\\
 $5799.20 \pm 0.88$ & -- & $1672 \pm 51$& --\\
 $674.6922 \pm 0.0098$ & -- & $684 \pm 8$ & $719 \pm 7$ \\
 -- & $584.4$ & -- & $516 \pm 4$ \\
\hline
\end{tabular}
\end{center}
\end{table}

\subsection{HIP12653 RV variation}
\label{RV}
\begin{figure}
\begin{center}
\includegraphics[width=0.8\linewidth]{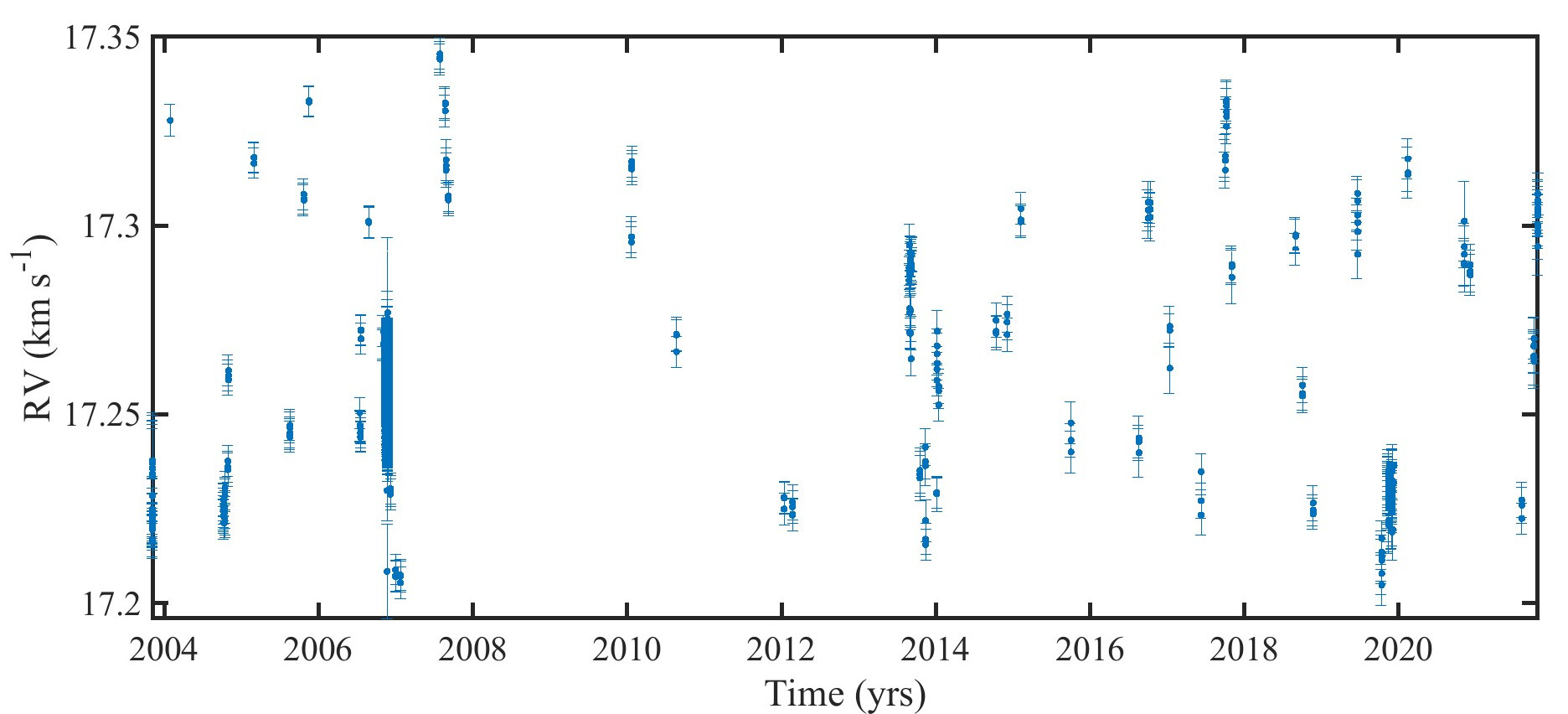}
\end{center}
\caption{RV curve of HIP12653.}
\label{fig:rvData}
\end{figure}

We applied the GLS periodogram to the RV timeseries shown in Fig. \ref{fig:rvData}, using the same methodology as described in Sect. \ref{method} for the S-index dataset. The planetary period corresponds to the highest peak in the GLS periodogram. Additional signals were identified by applying the GLS to the residuals after sequentially subtracting previously detected signals. A global fit including all previously identified periods was then performed and iteratively refined until no further significant periodicities remained. All orbital solutions were finally refined simultaneously. Orbtial parameters; (RV semi-amplitude, $K$, eccentricity, $e$, longitude of periastron, $\omega$, and time of passage through periastron, $T_0$), are also determined with the same method described for the S-index model.

The derived periods and corresponding orbital parameters are listed in Table \ref{tab:periodsRV}. We recovered the known planetary period at $307.755 \pm 0.049$ d. The signal at $3.77312 \pm 0.00014 $d likely corresponds to the first harmonic of the stellar rotation period (7.7 d), while the period of $678.1 \pm 1.8$ d is close to the possible secondary magnetic cycle period of $674.6922 \pm 0.0098$ d detected in the S-index. 

\begin{table*}
 \begin{center}
   \caption{\small Periods detected in the RV dataset along with the orbital parameters: semi-amplitude $K$, the eccentricity, $e$, longitude of periastron, $\omega$, and the time of passage through periastron $T_0$ (with reference epoch of $2454400$).}
    \vspace*{5mm}
    \label{tab:periodsRV}
    \begin{tabular}{rrrr}
    \hline
   & {$P_1 $} & {$P_2 $} & {$P_3 $} \\
\noalign{\smallskip}
\hline
\noalign{\smallskip}
P [d] & $307.755 \pm 0.049$ & $3.77312 \pm 0.00014$ & $678.1 \pm 1.8 $  \\
K & $57.94 \pm 0.69$  & $2.48 \pm 0.18$ & $8.09 \pm 0.65$ \\
e & $0.35806 \pm 0.00014$ & $0$ & $0$ \\
$\omega$ [deg] & $26.1670 \pm 0.0013$ & $0$ & $0$\\
$T_0$ [BJD] & $238.56 \pm 0.36$ & $0.529 \pm 0.048$ & $521.1 \pm 8.6$ \\
\hline
\end{tabular}
\end{center}
\end{table*}

\subsection{Exploring the data using photometry}

In order to further explore the existing rotational periods of HIP12653 we use photometric data. We will apply the multiharmonic fitting method to these data and cross-check the results with the findings presented in the previous sections. But first of all, we will proceed with testing the multiharmonic fitting method using Monte Carlo simulations of stars with ARs, as well as apply the method to actual observational Solar data. These steps are meant to explore the performance of multiharmonic fitting before applying it to HIP12653 data.    

\subsubsection{Testing the multi-harmonic method using simulated data}
\begin{figure}
\begin{center}
\includegraphics[width=0.8\linewidth]{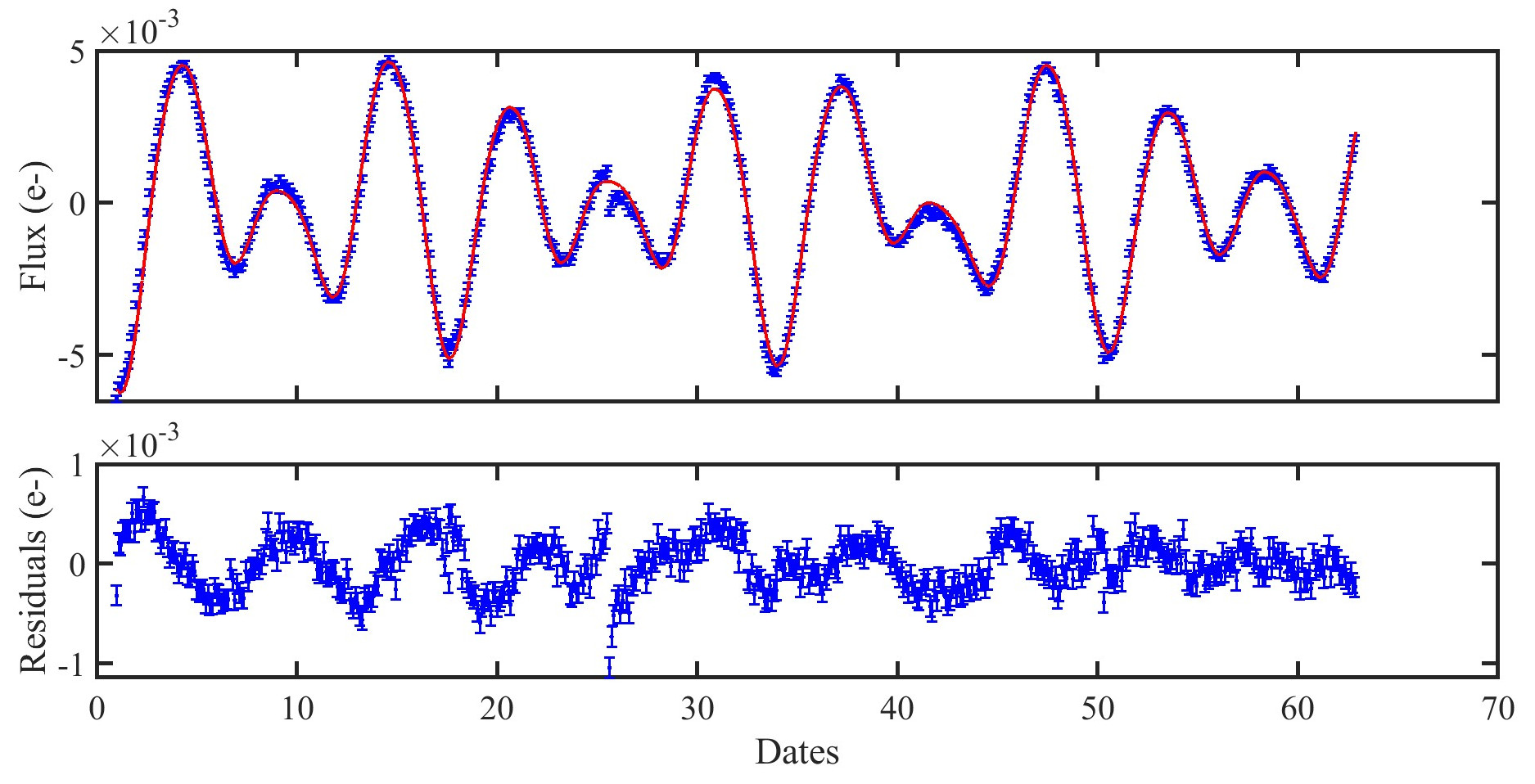}
\end{center}
\caption{Top panel: The simulated light curve (blue) is fitted with a multiharmonic sine function (red).
Bottom panel: Residuals after subtracting the best-fit model from the data. Some periodic variability remains, but it is not significant based on the RMSE criterion.}
\label{simulLC}
\end{figure}

We apply our multi-harmonic method to each of the $10000$ simulated light curves to estimate the possible rotation periods. Fig. \ref{simulLC} shows a simulated light curve with input periods of 8.65, 8.30, 5.51, and 10.20 d. Our method recovered three significant periodicities at 5.35, 8.24, and 10.81 d. While the input periods of 5.51 and 10.20-day are recovered within the frequency resolution, only one of the two closely spaced $\sim$8-day periods was recovered. This is not a failure of the method, but a consequence of the limited frequency resolution imposed by the time baseline of the simulation. With a 60-day observation window, the minimum resolvable frequency difference is $1/T \approx 0.017 d^{-1}$. The frequency separation between 8.65 and 8.30 d is only $\sim 0.005 d^{-1}$, below the resolution limit, making the signals effectively indistinguishable in the periodogram. The method can only resolve rotational harmonics if their periods are sufficiently separated, given the observational time span. The bottom panel of Fig.~\ref{simulLC} displays the residuals after subtracting the multi-harmonic fit. While these residuals exhibit low-amplitude systematic variations, a GLS analysis reveals a periodicity around 7 days, which does not correspond to any of the injected input periods. We interpret this signal as a likely alias or harmonic artifact arising from the limited frequency resolution of the 60-day observation window, rather than a missed rotational component. This is consistent with the simulation’s inability to resolve the two closely spaced $\sim$8-day signals, whose frequency separation falls below the resolution threshold of the dataset.

We estimate the rotational shear of the true periods from the maximum and minimum periods for each light curve as:
\begin{equation}
\alpha_{true} = \frac{P_{\rm{max}} - P_{\rm{min}}}{P_{\rm{max}}}  
\end{equation}
The corresponding estimated periods to ${P_{\rm{max}}}_{\rm{true}}$ and ${P_{\rm{min}}}_{\rm{true}}$ are used in calculating the estimated shear, $\alpha_{\rm{estimated}}$.

\begin{figure}
\begin{center}
\includegraphics[width=0.8\linewidth]{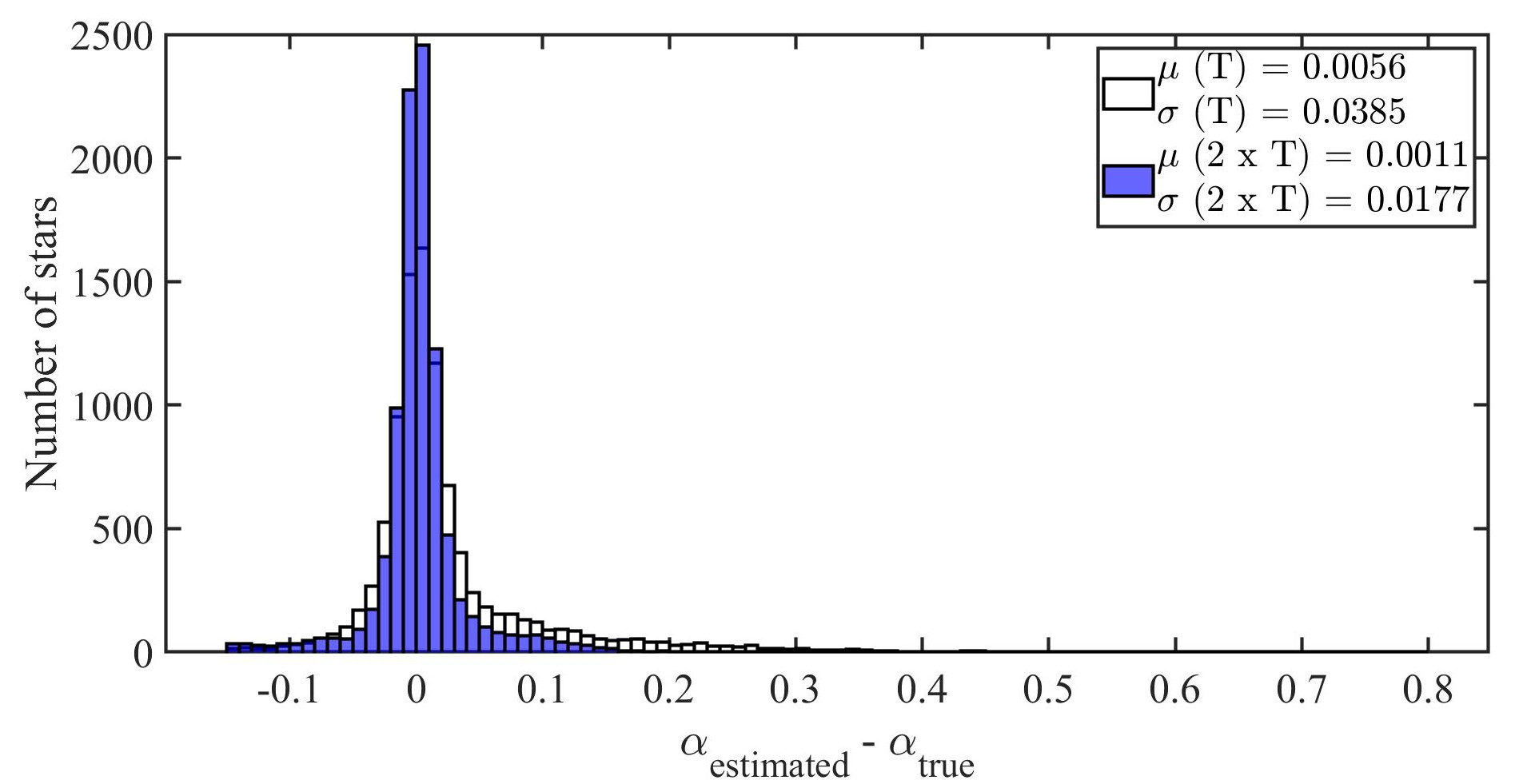}
\end{center}
\caption{Histogram of the difference between the true and estimated shear for simulations with a timespan around 30 d (white histogram) and simulations with a timespan around 60 d (blue histogram) and the same number of measurements.}
\label{alpha_soap}
\end{figure}

We show in Fig. \ref{alpha_soap} (white histogram) the distribution of the difference between the estimated and true shear values over the 30-day time span. The accuracy is $0.0056 \pm 0.0005$, indicating that, on average, the estimated shear closely matches the true shear with a standard deviation of $0.0385 \pm 0.0005$. Notably, the distribution exhibits a positive skew, with a longer tail extending towards positive values. This skew suggests that, while most estimated shear values are accurate, the method occasionally overestimates the shear. The tail and sparse bins beyond 0.10 highlight a few cases of significant overestimation. The asymmetry is likely due to difficulties in detecting closely spaced rotation periods, which is expected as the peaks separation in the periodogram is limited by the frequency resolution $\Delta f = 1/ T$,  with $T$ being the simulation timespan.  To investigate this further, we used simulations with doubled timespan while maintaining the same number of measurements. The resulting distribution is overplotted in Fig. \ref{alpha_soap} (blue histogram). With the longer timespan, the accuracy improves to $0.0011 \pm 0.0003$ and a standard deviation decreases to $0.0177 \pm 0.0002$. The reduced tail and improved resolution of closely spaced periods demonstrate the benefits of a longer observation window. This is true if the lifetime of the ARs is long enough.

For larger true shears, the differential rotation signal is stronger, enabling more accurate estimates regardless of the timespan.

To evaluate the accuracy of our estimated periods, we compute the ratio of periods for which the difference between the true and estimated values is within three times the period uncertainties, $|P_{\rm{estimated}} - P_{\rm{true}}| < 3 \sigma_P$, relative to the total number of periods. Rejecting periods with uncertainties larger than a third of the estimated periods, the success rate is $74.2\%$, with the fraction of stars with input periods that do not have a corresponding match among the estimated periods is $17.2\%$. 

\begin{figure}
\begin{center}
\includegraphics[width=0.8\linewidth]{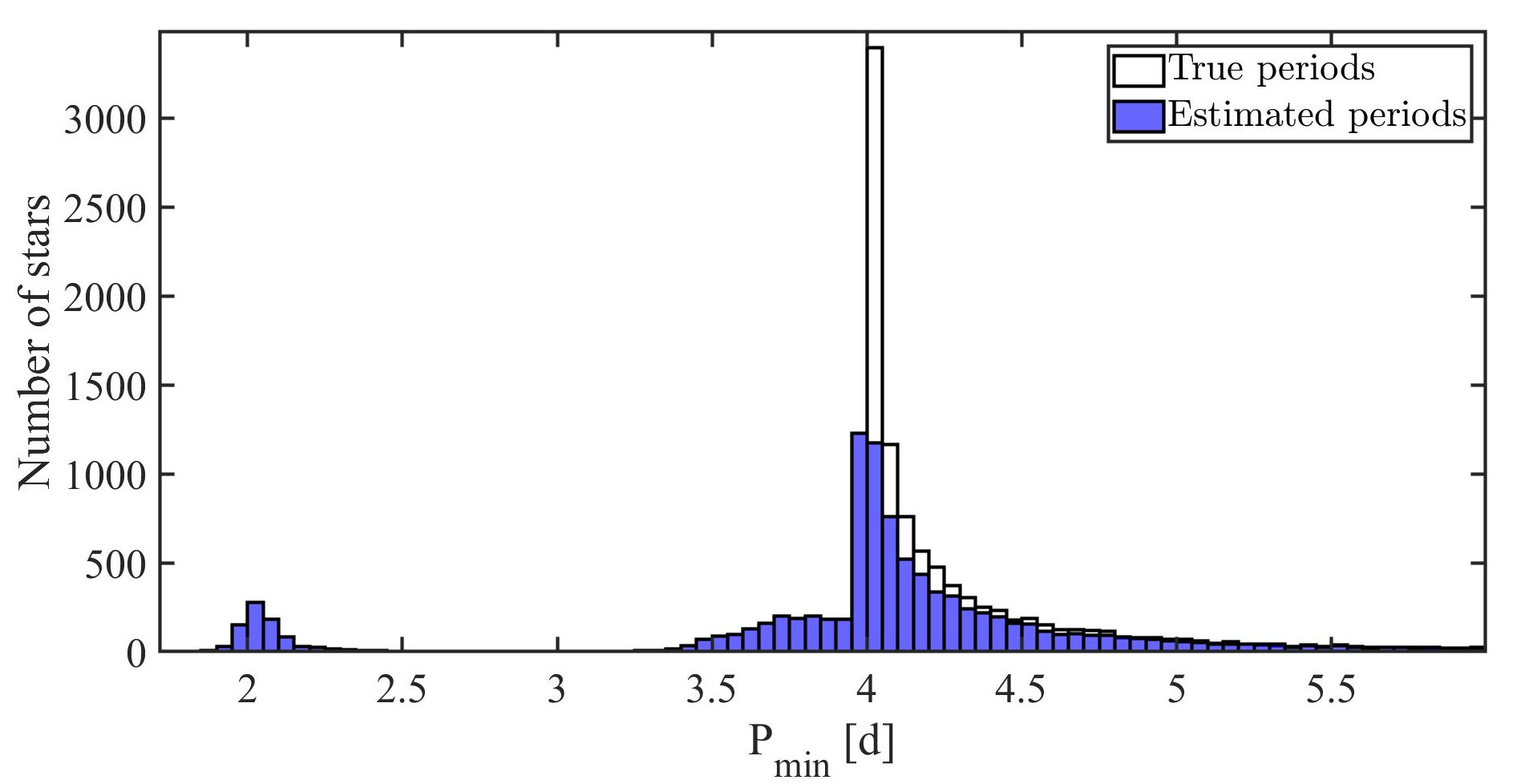}
\end{center}
\caption{Histogram of lowest true periods (white) and estimated periods (blue).}
\label{lowP_soap}
\end{figure}

To check whether we can detect periods shorter than the equatorial period, we present in Fig. \ref{lowP_soap} a histogram of the lowest true periods (in white) and estimated periods (in blue). The distribution of the lowest estimated periods is multimodal, with a primary peak around 4 days, as expected, since the true periods also cluster around this value. A secondary peak appears near 2 days, likely due to the presence of the first harmonic in the signal, which is expected when ARs are separated by $180^{\circ}$-longitude. This scenario accounts for only $8.46\%$ of the sample. Cases where the lowest period falls below the equatorial period of 4 days and where $|P_{\rm{estimated}} - P_{\rm{true}}| > 3 \sigma_P$ make up $12.7\%$ of the sample.

To explain these 'missed' minimum periods, we compare the frequency difference between two close periods, $|\frac{1}{P_1} - \frac{1}{P_2}|$, to the frequency resolution. We find that in $65\%$ of cases (within the $12.7\%$ subset), the periods cannot be distinguished in the periodogram because their frequency difference is smaller than the resolution limit, i.e., $|\frac{1}{P_1} - \frac{1}{P_2}| < \frac{1}{\Delta T}$. As a result, the periodogram merges these periods into a single, blended peak, preventing their individual identification. Consequently, only $4.45\%$ of the whole sample have unexplained lowest estimated periods.

Given a minimum of two ARs and the objective of determining the differential rotation, we define the success rate as the fraction of iterations in which at least two estimated periods fall within three times the uncertainties relative to the total number of iterations. Using our multi-harmonic method, we achieve a success rate of $92.1\%$, which exceeds the $64\%$ success rate reported by \cite{reinhold2013fast}. Their approach, based on a single-harmonic method, identified two distinct periods by selecting the second period according to maximum shear.

\subsubsection{Applying multiharmonic fitting to Solar data}

Given that the butterfly diagram of the solar cycle is well-established, we apply our periodogram analysis to the total solar irradiance (TSI) data collected by the VIRGO experiment \citep{finsterle2021total} aboard the SoHO satellite \citep{frohlich1995virgo}. We utilize Level 2 data, which includes measurements taken in a bolometric passband with a daily cadence, obtained from the official webpage\footnote{\url{https://www.pmodwrc.ch/en/research-development/solar-physics/tsi-composite/}}.

\begin{figure}[h]
\begin{center}
\subfloat{\includegraphics[width=0.6\linewidth, trim={0 2cm 0 0},clip]{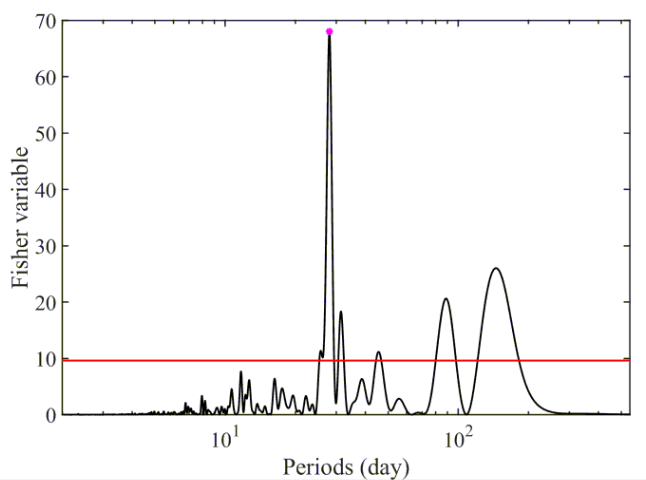}}\\
\subfloat{\includegraphics[width=0.6\linewidth, trim={0 2cm 0 0},clip]{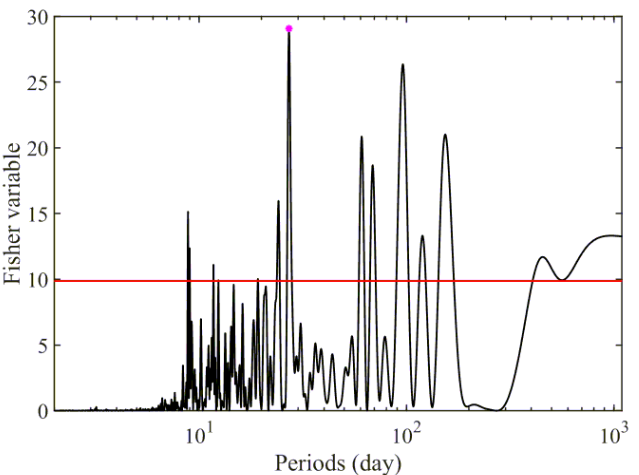}}\\
\subfloat{\includegraphics[width=0.6\linewidth]{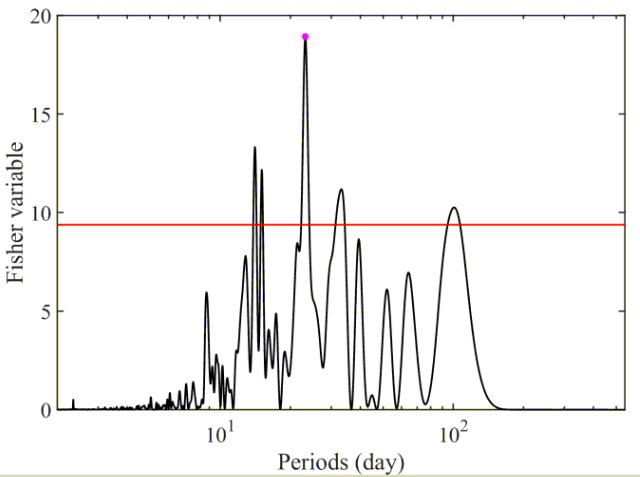}}
\end{center}
\caption{GLS periodogram applied to the Sun's total irradiance. Top panel: $2009-2010$ dataset, where the highest peak occurs at a period of 28 d. Middle panel: $2013-2014$ dataset, with the highest peak at 27.1 d. Bottom panel: $2019-2021$ dataset, where the highest peak is at 23.3 d.}
\label{cycleSun}
\end{figure}

Solar Cycle 24 started in December 2008, reached its peak in April 2014, and ended in December 2019. To analyse this, we utilize three datasets that correspond to the beginning, peak, and end of the cycle, with the following timeframes: $2009-2010$, $2013-2014$, and $2019-2020$. We show in Fig. \ref{cycleSun} the three periodograms corresponding to each dataset. At the beginning of the cycle (first minimum activity), we determine a period of $28.007 \pm 0.013$ d (top panel), which likely corresponds to sunspots at high latitudes. At the cycle's maximum, we find a period of $27.1106 \pm 0.0015$ d (middle panel), followed by another period of $24.2069 \pm 0.0014$ d, corresponding to sunspots at mid latitudes. By the end of the cycle, we determine a period of $23.204 \pm 0.013$ d (bottom panel), which is associated with sunspots near the equator (second minimum activity). These varying periods clearly reflect the butterfly diagram, illustrating the Sun’s differential rotation.

These results are consistent with previous studies of Solar Cycle 24. \citet{wauters2016lyra} analysed LYRA channels, sunspot number, sunspot area, and flare index between 2010 and 2014, applying both the Lomb–Scargle periodogram and wavelet transform. They reported dominant periodicities ranging from 24 days (in sunspot area) to 28 days (in flares), which match well with our detections of 24.2 d during 2013–2014 and 28 d during 2009–2010. Similarly, \citet{sharma2021differential} analyzed solar full-disk images at 30.4 nm from the Extreme Ultraviolet Imager (EUVI) onboard the STEREO spacecraft, covering the period $2008-2018$. They measured rotation rates that varied throughout the cycle: around 14.97 deg/day ($\sim 24d$) near activity maximum in 2014, gradually decreasing to 13.9 deg/day ($\sim 26$ d) during the ascending and declining phases, reaching $\sim 28$ d at high latitudes in 2010.

Applying our multi-harmonic method to solar irradiance reveals the presence of different rotation periods at various phases of the magnetic cycle. 

\citep{lanza2014measuring} applied a method aimed at detecting differential rotation from evenly-sampled photometry by modelling light curves with a two-spot analytical model and using Bayesian model selection. However, when applied to the Sun-as-a-star data (TSI photometry over a 200-day interval starting from 30 November 1996), their method failed to recover the true periods. This is primarily because the Sun’s photospheric active regions evolve rapidly, while their approach assumes the presence of long-lived, stable surface features that produce coherent rotational modulation over several cycles. Additionally, their method requires prior knowledge of the stellar rotation axis inclination to define the \textit{a priori} distribution of the inclination, which is essential for convergence.

In contrast, by employing a multi-harmonic method that retrieves periods based on signal RMSE, independent of unknown parameters such as inclination, we achieved a reliable estimate of differential rotation in both simulated data and the Sun. This demonstrates the effectiveness of our approach. However, a limitation of our method is that ARs have specific lifetimes and may not complete enough rotations within the observation period, preventing their period from appearing in the periodogram.

\subsubsection{HIP12653 TESS photometry}
\label{tess}
\cite{katsova2010differential} found that seasonal variations in stellar rotation rate are due to latitude variation of the activity belt according to the phase of its cycle, since they noticed that slower rotation occurs during epochs of high activity for their target star.

For a differentially rotating stellar surface, a minor variation in the latitude where spots emerge leads to a faint, quasi-periodic fluctuation in the photometric rotation period throughout the activity cycle \citep{vida2014looking}. Due to differential rotation, spots emerge at varying latitudes, and as the activity belt shifts in the butterfly diagram, multiple peaks appear in the periodogram. This can also lead to changes in the observed main rotation period.

In order to confirm (or otherwise) the differential rotation of HIP12653, we analysed the star's light curve obtained by the TESS mission across multiple years. The data were retrieved from Mikulski Archive for Space Telescopes (MAST\footnote{\url{https://mast.stsci.edu/portal/Mashup/Clients/Mast/Portal.html}}) using LightKurve \citep{2018ascl.soft12013L}. In 2018, TESS observed the star in two consecutive sectors (S2 and S3) with an exposure time of 1800 seconds. In 2020, it was observed again in sectors S29 and S30 with a shorter exposure time of 600 seconds. Most recently, in 2023, the star was observed in sector 69 with an exposure time of 200 seconds. 

We apply our multiharmonic fitting method to these three datasets, stitched and normalised, separately, as they might correspond to different phases of the magnetic cycle. We also tested the software Period04 \citep{lenz2005period04}, which is based on discrete Fourier transformation. The results obtained were fully consistent with those derived using the method adopted in this study, and therefore, no further details are reported. We list in Table \ref{tab:tess} the periods selected for each dataset; we exclude periods close to the duration of the data gaps. 

\begin{figure}
\begin{center}
\subfloat{\includegraphics[width=0.6\linewidth, trim={0 2.0cm 0 0},clip]{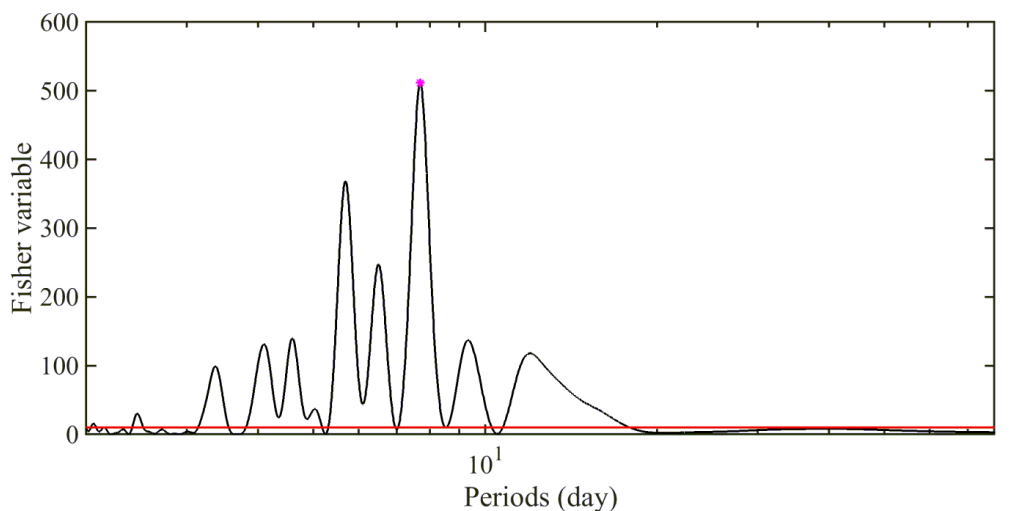}}\\
\subfloat{\includegraphics[width=0.625\linewidth, height=5cm]{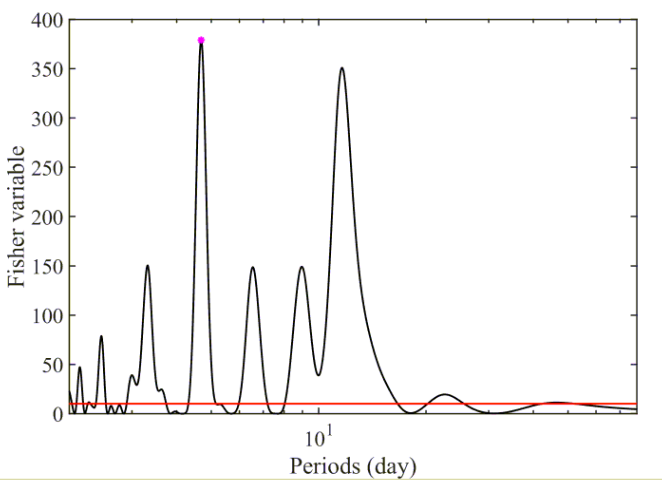}}
\end{center}
\caption{GLS periodogram applied to TESS photometric dataset of sectors S2 and S3.}
\label{period18_1}
\end{figure}

\begin{figure}
\begin{center}
\includegraphics[width=0.8\linewidth]{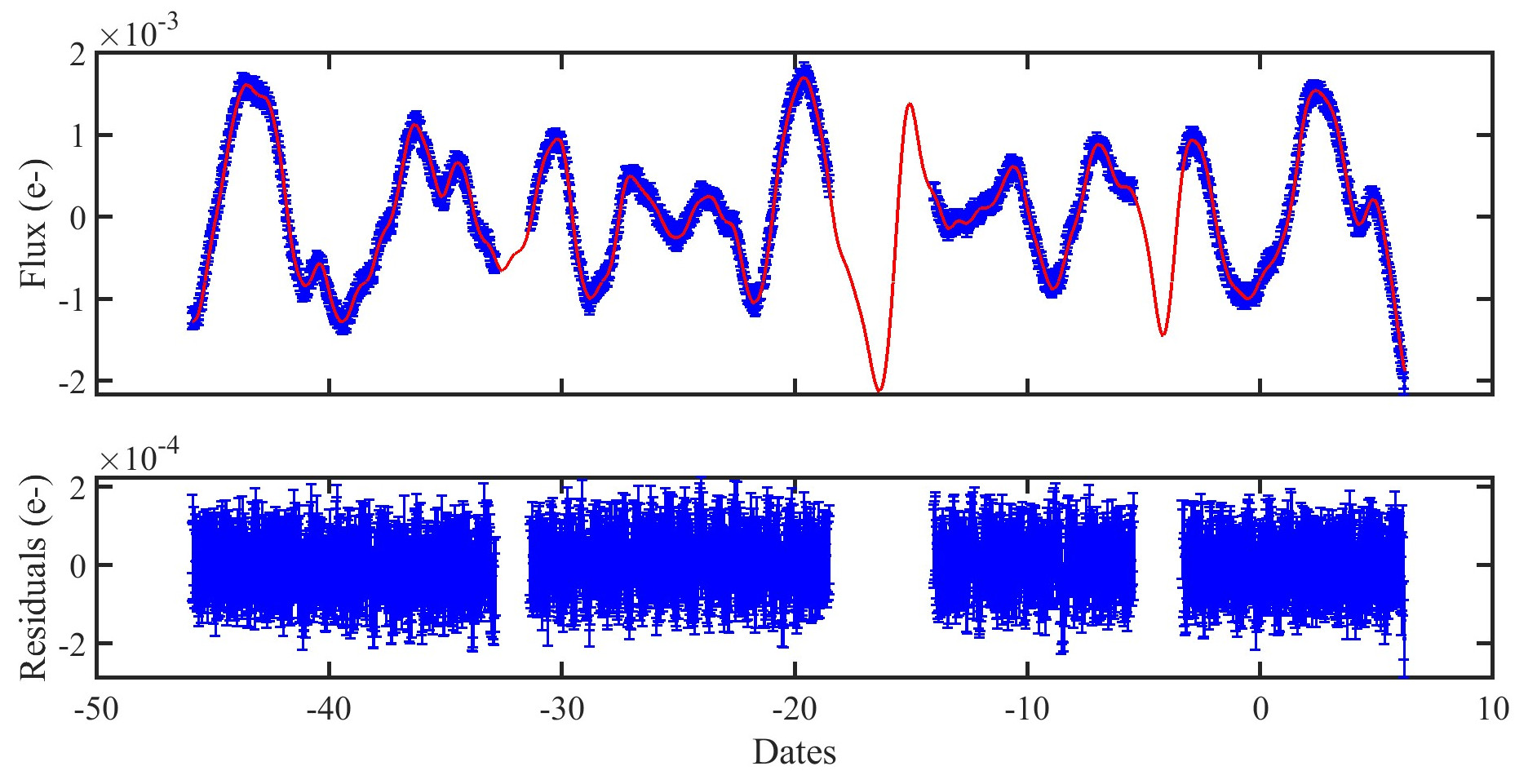}
\end{center}
\caption{Top panel: Sectors S2 and S3 light curve in blue with the best fit overplotted in red. Bottom panel: Residuals after subtracting the best-fit model from the data.}
\label{fig:tess18}
\end{figure}

In sectors S2 and S3, the periodogram initially shows three dominant peaks at 7.7 d, 5.6 d, and 6.5 d (top panel of Fig. \ref{period18_1}). After removing the 7.7-day period, the 5.6-day peak becomes more pronounced, suggesting that it is independent and possibly linked to another AR. On the other hand, the peak at 6.5 d is significantly reduced in the third periodogram, and we see the emergence of a peak at 4.8 d (bottom panel of Fig. \ref{period18_1}). We show in Fig. \ref{fig:tess18} the light curve in blue with overplotted fitted curve in red.  

In our simulations, we found that only $4.45\%$ of the sample have unexplained lowest periods, while $8.25\%$ are due to the fact that the periods are unresolved. If we consider the two periods of $7.7$ d and $5.6$ d to be independent, their frequency difference is 0.05, which is larger than the frequency resolution of 0.02, it is therefore unlikely for the 4.8 d to be an alias.

Additionally, according to \cite{strassmeier2009starspots}, magnetic features tend to concentrate towards the polar regions in highly active stars. Since our target is an actively intermediate star, more active than the Sun, the spots are expected to be concentrated at latitudes higher than the Sun's activity belt of $\pm 30^{\circ}$. This suggests that the 7.7-day period corresponds to a region far from the equator.

Furthermore, the detection of two distinct periods during the activity maximum of the Sun (24 d and 27 d) suggests that the multiple periods observed in this TESS run of HIP12653 correspond to rotation at different latitudes.

\begin{figure}
\begin{center}
\includegraphics[width=0.8\linewidth]{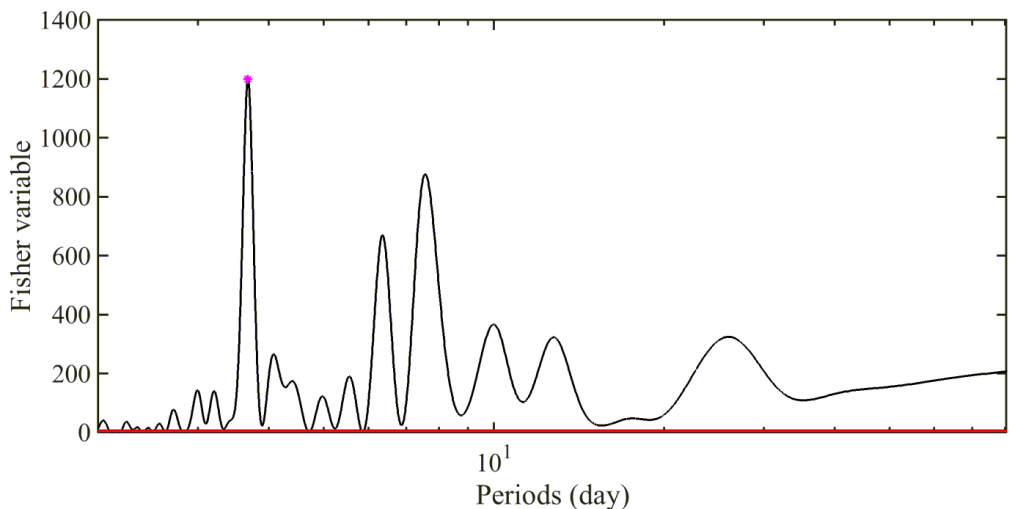}
\end{center}
\caption{GLS periodogram applied to TESS photometric dataset of sectors S29 and S30.}
\label{period20_1}
\end{figure}

In sectors S29 and S30, the first periodogram (Fig. \ref{period20_1}) reveals a strong peak at 3.7 days, followed by another at approximately 7.5 days. Once the 3.7-day solution is subtracted, the 7.5-day peak diminishes, indicating that the 3.7-day period is likely a harmonic of the 7.5-day period. Consequently, the 7.5-day period was selected as the primary rotation period. These two sectors exhibit a higher noise level compared to the previous ones. To account for this, we analysed each 14-day segment separately and identified the periods listed in the second column of Table \ref{tab:tess}, excluding repeated values across the four subsets. Compared to sectors S2 and S3, the detected periods are remarkably similar, which is expected if these sectors correspond to the same phase of the magnetic cycle, assuming a period of $674.7$ days.

Similarly, sector S69 exhibits an even higher noise level, making it difficult to retrieve the exact periods. The 3.7-day period may correspond to the first harmonic of the 7.7-day period, while the 9.1-day period could be twice the 4.5-day period (subharmonic) or a period of an AR at a very high latitude. This suggests that between 2018 and 2023, the rotation period varied from a minimum of 4.8 days near the equator to 7.7 days at higher latitudes.

 \begin{table}
 \begin{center}
   \caption{\small Periods detected in the TESS photometric dataset.}
    \vspace*{5mm}
    \label{tab:tess}
    \begin{tabular}{rrrr}
    \hline
    &{Sectors S2 S3} & {Sectors S29 S30} & {Sectors S69} \\
\noalign{\smallskip}
\hline
\noalign{\smallskip}
$P_1$ [d] & $7.687 \pm 0.055$ & $7.282 \pm 0.019$ & $9.0586 \pm 0.0051$  \\
$P_2$ [d] & $5.650 \pm 0.045$ & $2.523 \pm 0.083$ & $3.7033 \pm 0.0026$  \\
$P_3$ [d] & $4.802 \pm 0.074$& $4.45 \pm 0.26$& \\
\hline
\end{tabular}
\end{center}
\end{table}

We compare in Table. \ref{tab:TessSindex} rotation periods determined from spectroscopic data (including results from all data, and those with minimum activity measurements excluded) and different TESS sectors.
\begin{table}
 \begin{center}
   \caption{\small Comparison of periods detected in the S-index and TESS photometric datasets.}
    \vspace*{5mm}
    \label{tab:TessSindex}
    \begin{tabular}{ccc}
    \hline
    &{S-index} & {TESS}  \\
\noalign{\smallskip}
\hline
\noalign{\smallskip}
$P_1$ [d] & $4.7699949 \pm 0.0000021$ & $4.802 \pm 0.074$ \\
$P_2$ [d] & $7.6383684 \pm 0.0000089$ & $7.687 \pm 0.055$ \\
$P_3$ [d] & $3.7536551 \pm 0.0000020$ & $3.7033 \pm 0.0026$\\
$P_4$ [d] & -- & $5.650 \pm 0.045$\\
\hline
\end{tabular}
\end{center}
\end{table}
The presence of the $4.8$ d period in the TESS data confirms that the period detected in the S-index data ($4.77$ d) is the rotation period of ARs at a different latitude.

\begin{figure}
\begin{center}
\subfloat{\includegraphics[width=0.6\linewidth, trim={0 3cm 0 0},clip]{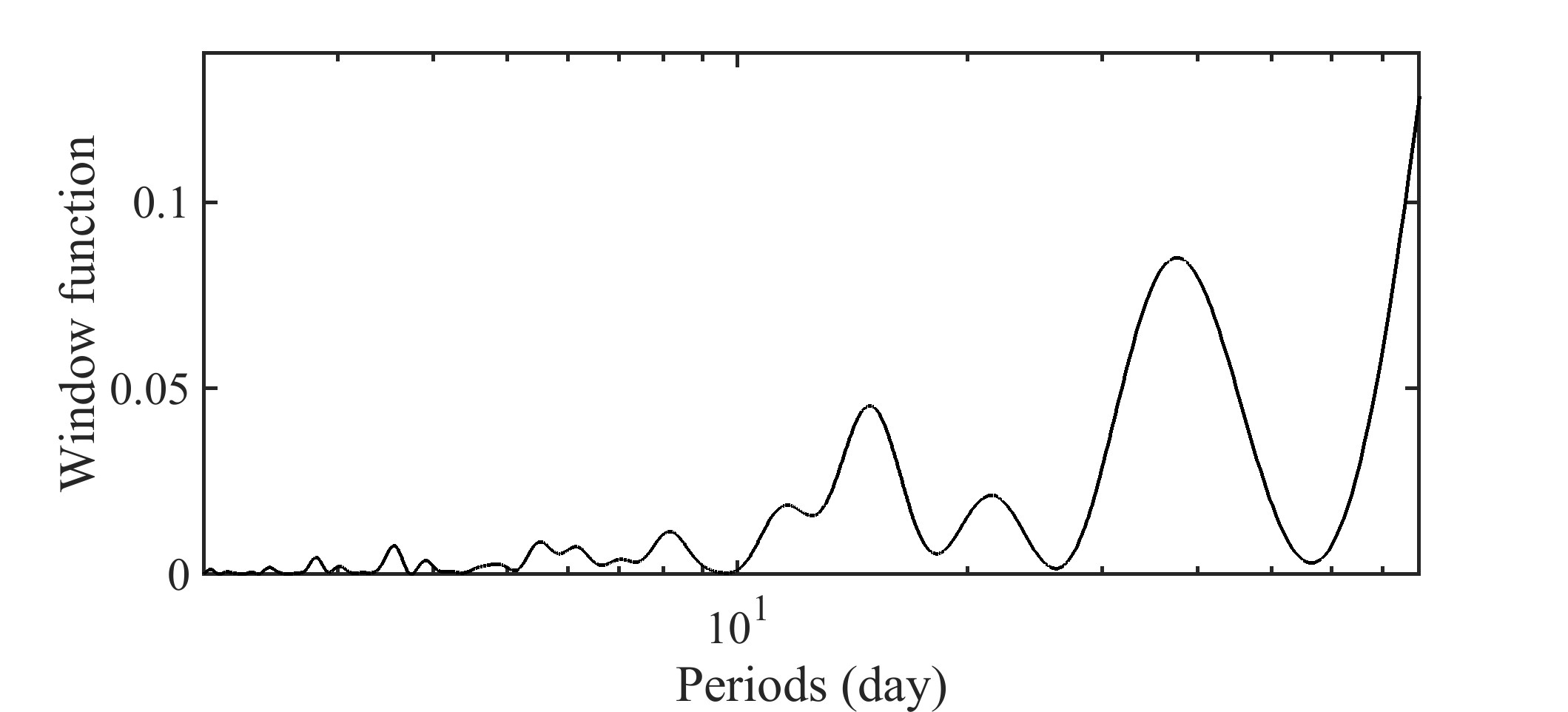}}\\
\subfloat{\includegraphics[width=0.6\linewidth, trim={0 3cm 0 0},clip]{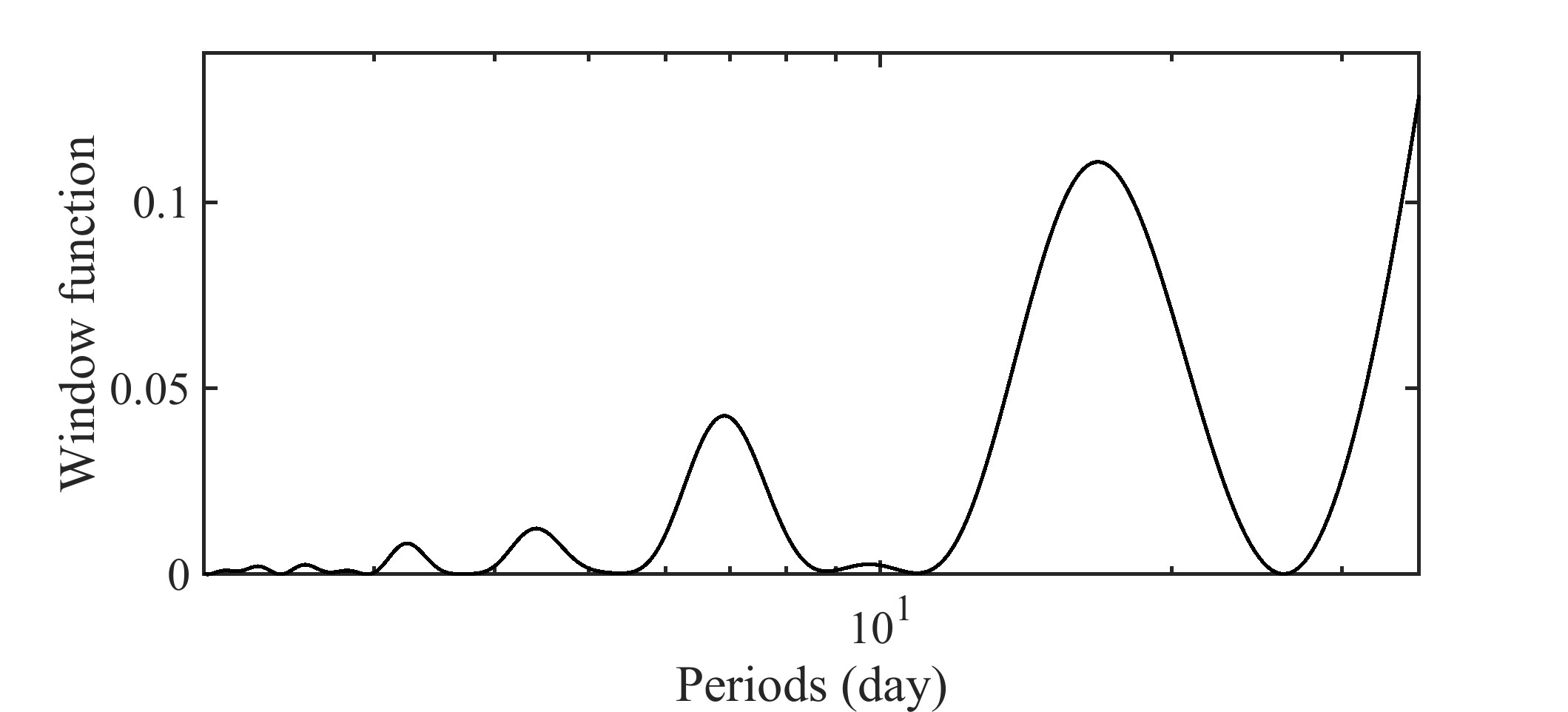}}\\
\subfloat{\includegraphics[width=0.6\linewidth]{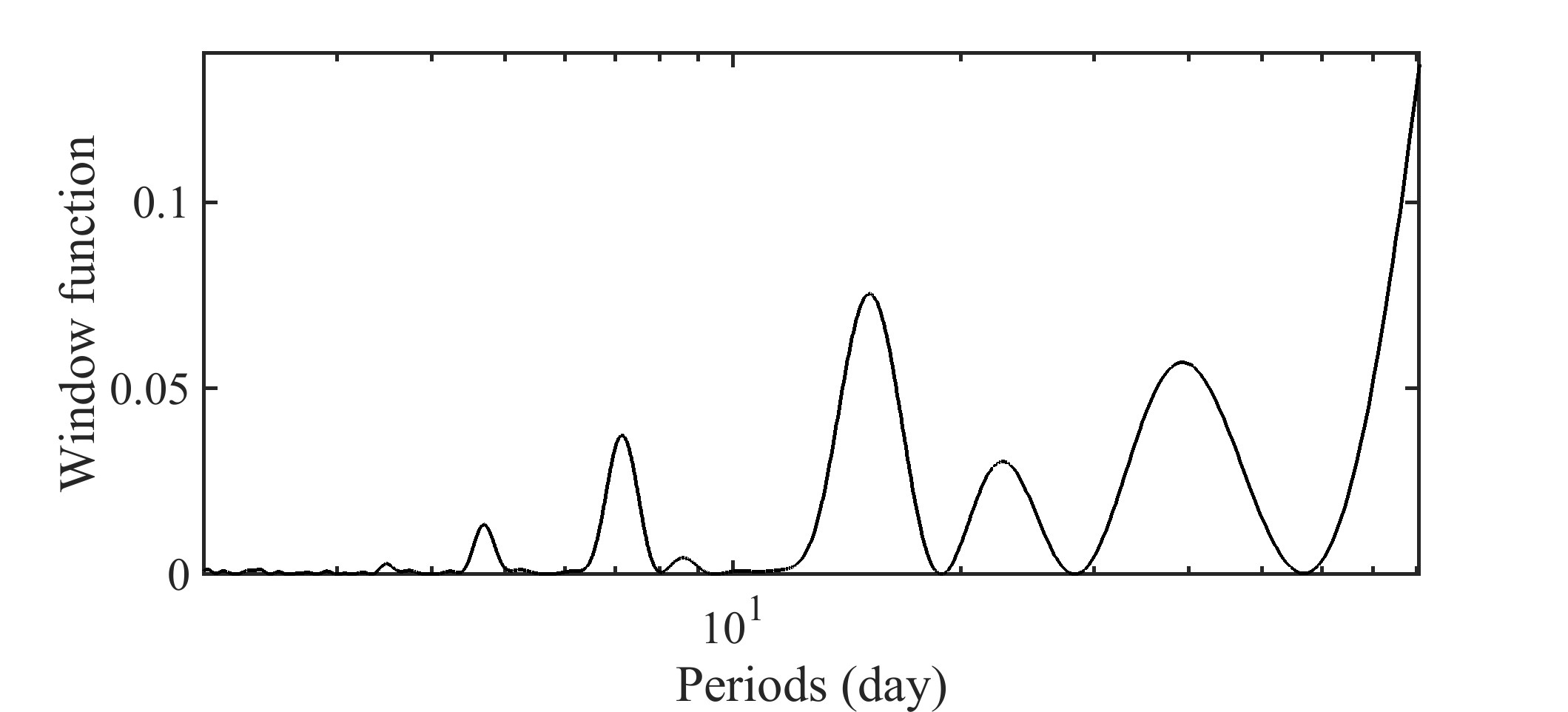}}
\end{center}
\caption{Window function of TESS data for sectors S2 and S3 (top panel), S29 and S30 (middle panel), and S69 (bottom panel).}
\label{fig:WF}
\end{figure}

To evaluate the potential influence of the sampling pattern on the detected periods, we computed the window function for each TESS dataset, as shown in Fig.~\ref{fig:WF}. The window function of the first dataset (Sectors 2 and 3) exhibits broad peaks at long periods ($>10$ days), but with amplitudes not exceeding $15\%$, suggesting that the detected GLS periods in this dataset are unlikely to be caused by sampling aliases. For the second dataset (Sectors 29 and 30), the window function also shows a main peak at periods longer than 10 days. One of the detected GLS periods ($7.28$ d) lies near a minor feature in the window function, but this feature is of low amplitude ($< 5\%$) and does not correspond to a main lobe. This supports the interpretation of the $7.28$~d signal as astrophysical. The window function of the third dataset is broader, with a pronounced feature around 15–20 days, likely reflecting the observational baseline and duration of the sector.

\subsubsection{Wavelet analysis}
To confirm the periods detected in photometric data with our method, we performed a wavelet transform to analyze the temporal evolution of the signal. Since our data contain gaps, we used the \textit{Weighted Wavelet Z-transform} (WWZ) method, which is well-suited for unevenly sampled time series. The analysis was carried out using the \texttt{wwz} Python package \citep{2023ascl.soft10003K}, which is based on the algorithm developed by \citet{foster1996wavelets}.

\begin{figure}
\begin{center}
\subfloat{\includegraphics[width=0.8\linewidth, trim={0 0.7cm 0 0},clip]{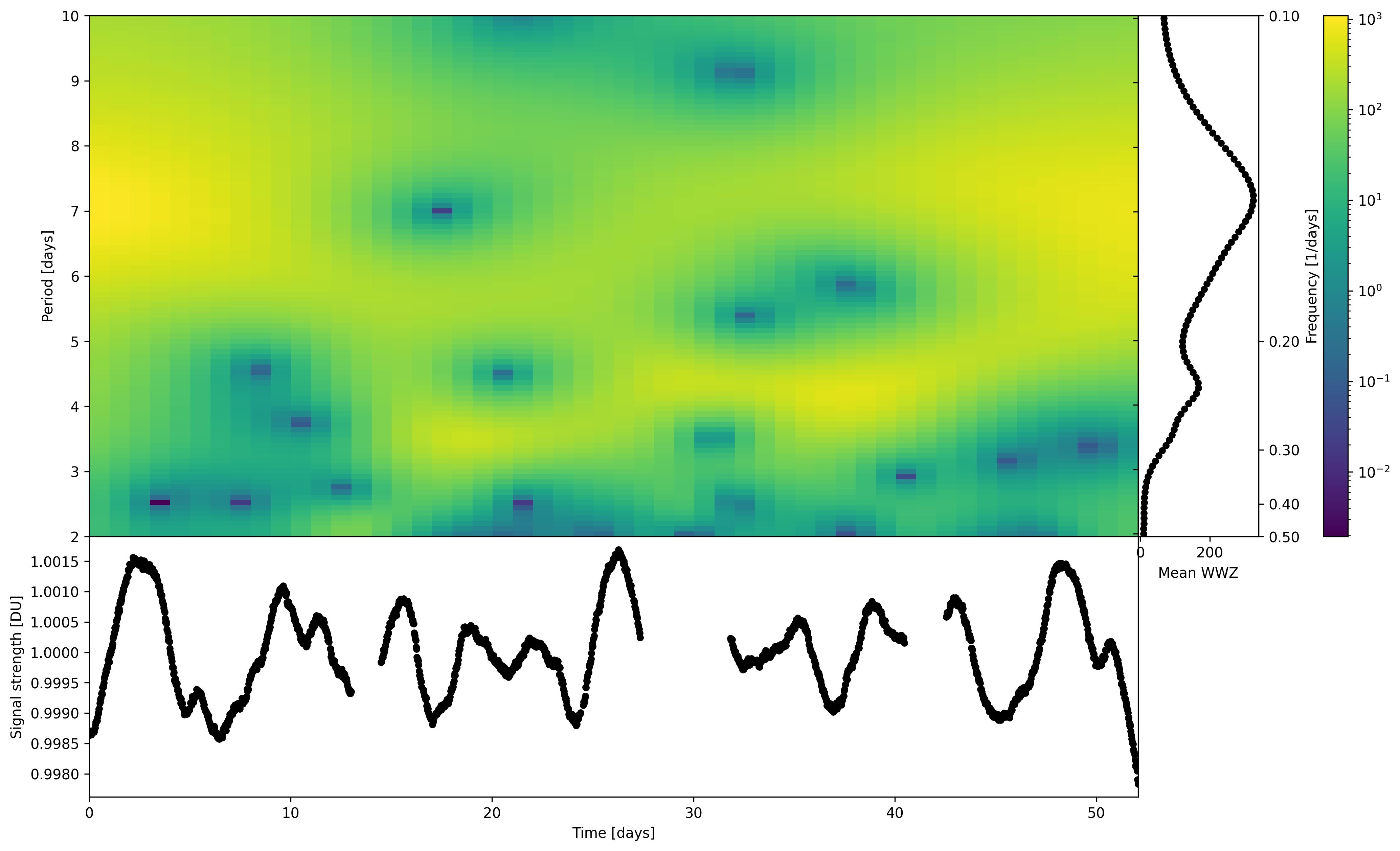}}\\
\subfloat{\includegraphics[width=0.8\linewidth, trim={0 0.7cm 0 0},clip]{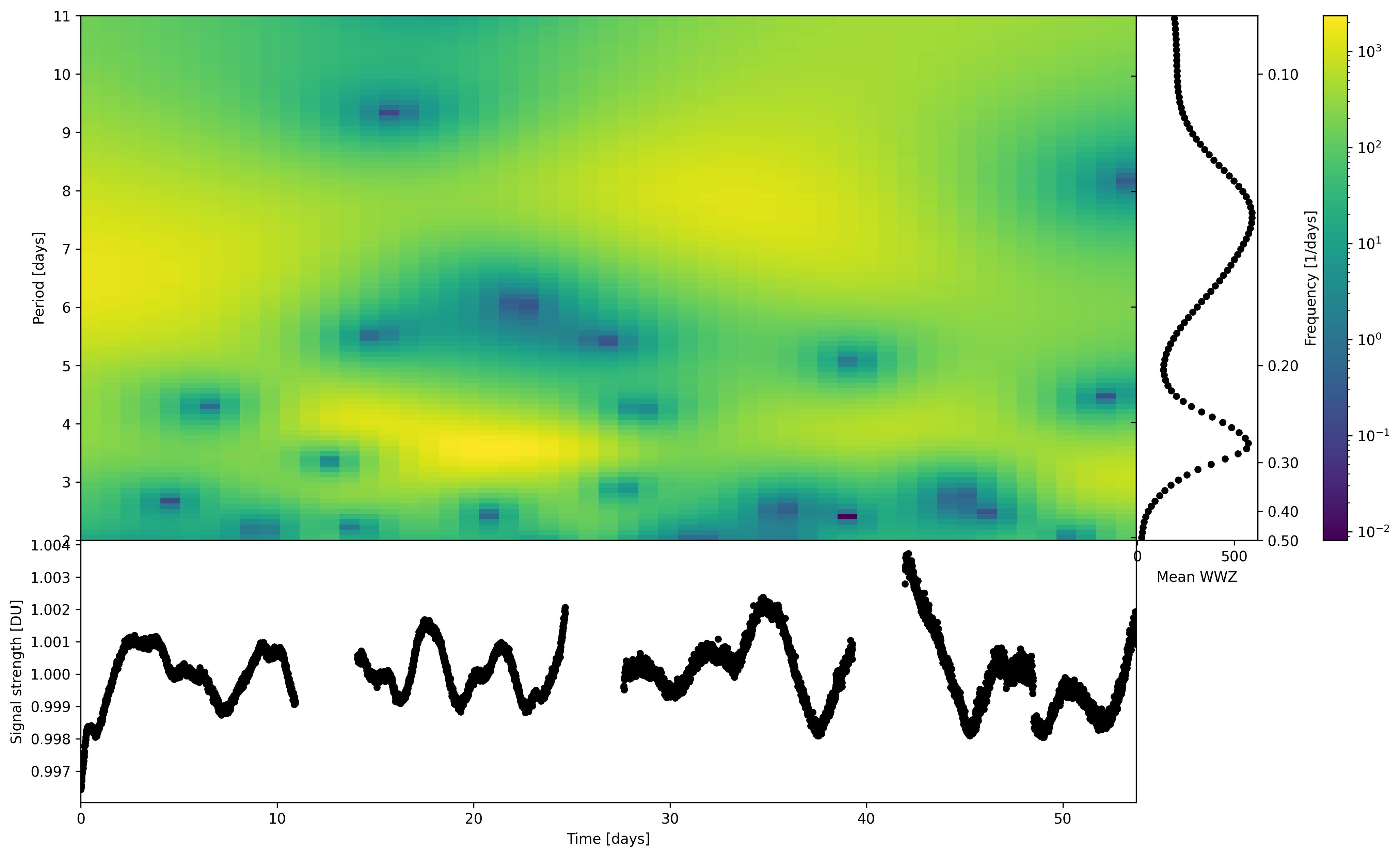}}\\
\subfloat{\includegraphics[width=0.8\linewidth]{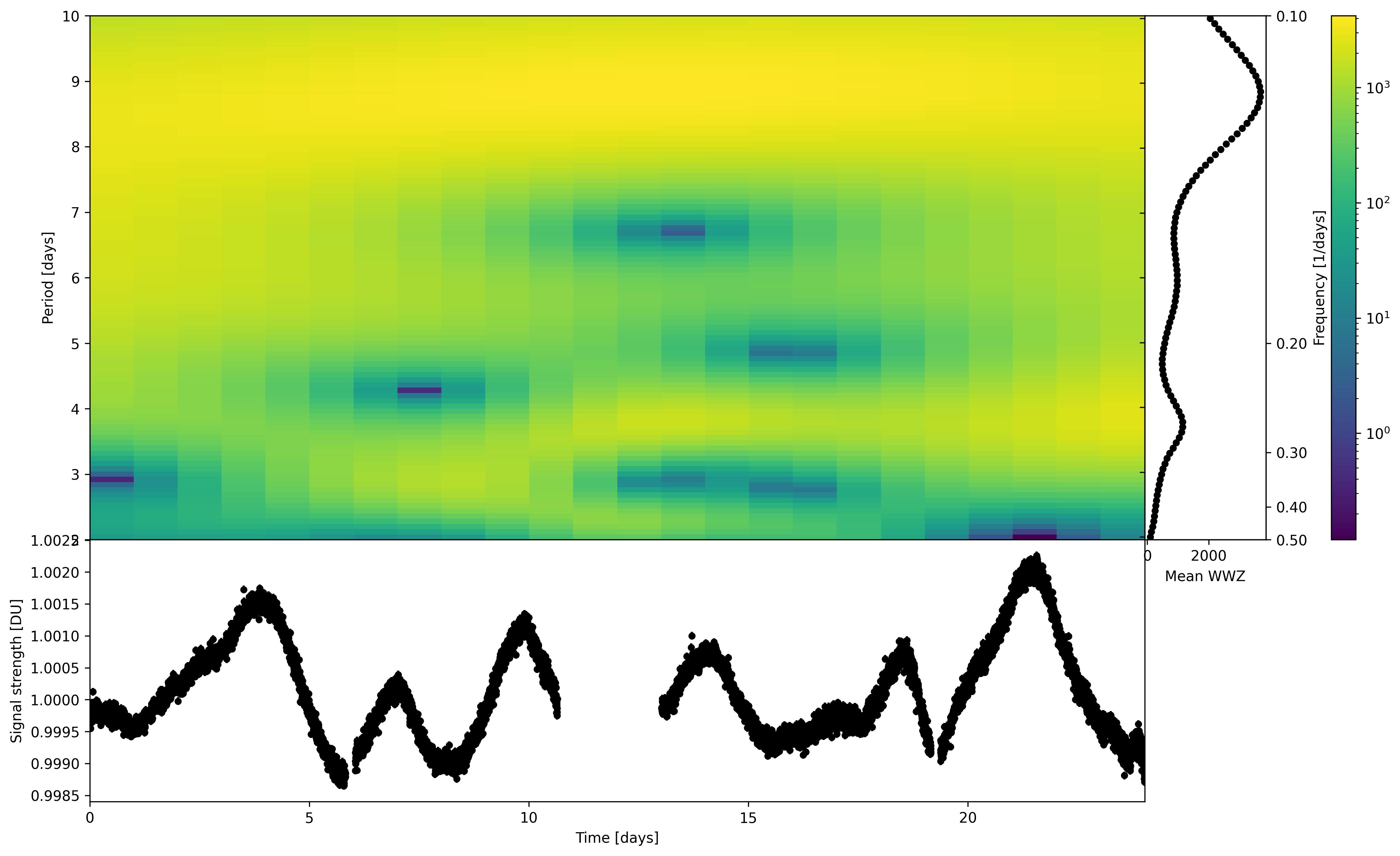}}
\end{center}
\caption{Local wavelet maps of the brightness variations of HIP12653 (center panels). The corresponding light curves are shown below each map, and the global wavelet spectra are shown to the right. Top: Sectors 2 and 3; Middle: Sectors 29 and 30; Bottom: Sector 69.}
\label{fig:Wavelet}
\end{figure}

The wavelet maps and their corresponding global spectra are shown in Fig. \ref{fig:Wavelet} for the three dataset. In each panel, the time series is shown at the bottom, the local wavelet power spectrum is displayed in the center as a color contour map, and the time-averaged (global) spectrum is shown on the right. In the top panel, which displays data from Sectors S2 and S3, two dominant periodicities are evident in the local map: the most significant, a $\sim$7-day period, persists during the first and last ten days, while a $\sim$4-day period emerges between days 30 and 45. Toward the end of the time series, a broad ridge spans periods from 5 to 9 days. All three periods detected in the GLS periodogram ($7.69$, $5.65$, and $4.80$~d) are clearly represented in the wavelet map. In the middle panel, corresponding to Sectors 29 and 30, three ridges of enhanced power are observed: one centered around 6.6 days at the beginning of the dataset, a second extending from 3 to 4.5 days throughout the full time span but with lower power, and a third centered around 7.9 days, dominating most of Sector 30. The bottom panel, showing Sector 69, reveals a persistent dominant period of approximately 8.9 days across the entire time span, along with a secondary signal around 3.7 days. Both periods are consistent with those detected in the GLS periodogram.

The wavelet maps reveal evolving periodicities in the TESS light curves that are consistent with the GLS and multiharmonic results, thereby providing additional support for the astrophysical origin of the detected signals.

\subsubsection{Rotation shear}
Differential rotation can be estimated from Doppler-broadened line profiles using the Fourier zero-ratio method ($q_2/q_1$), as proposed by \citet{reiners2002feasibility} and \citet{ammler2012new}. However, our target has $v \sin i = 6.84 \pm 0.36 ,\mathrm{km,s^{-1}}$, which is below the recommended threshold ($\sim$10 km s$^{-1}$) for a reliable application of this technique. No published study has successfully measured differential rotation with $v \sin i < 10$ km s$^{-1}$ using this method. Simulations by \citet{takeda2020detection} show that below this limit, the positions of the Fourier zeros deviate significantly and become unreliable, while empirical surveys by \citet{reiners2002feasibility} and \citet{ammler2012new} confirm that $\alpha$ measurements are trustworthy only for $v \sin i \gtrsim 12$ km s$^{-1}$ (CES) or $\gtrsim 45$ km s$^{-1}$ (FEROS). Although the large number of HARPS spectra available improves the signal-to-noise ratio, the reliability of Fourier analysis at this rotational velocity remains limited. Therefore, since our target lies outside the validated range of the method, we instead rely on photometric signatures of differential rotation.

Our method does not provide information on the specific latitude associated with the short period, nor does it allow us to determine whether the differential rotation is solar-like or anti-solar. However, following the statements by \cite{kuker2005differential} and \cite{lanza2014measuring} that main-sequence F-, G-, and K-type stars exhibit solar-like differential rotation, the 4.8 d period could correspond to the rotation period near the equator. Using the projected rotational velocity vsini of $6.84\, km s^{-1}$ obtained in this work and the radius of $1.16 \pm 0.04 \,R_{\odot}$ \citep{bruntt2010accurate}, and a rotation period of 4.8 days, which we interpret as the equatorial rotation period, we obtain an inclination of $34.0 \pm 1.8 ^{\circ}$. In contrast, if we adopt the longer period of 7.7 d, the inferred inclination becomes $63.9 \pm 2.2 ^{\circ}$.
Assuming the period at the highest latitudes is $7.7$ d, we derive a lower limit of the relative differential rotation shear of the activity belt of $\alpha = 0.38 \pm 0.01$, and a horizontal shear of $\Delta\Omega = 0.49\pm 0.11\, \rm{rad\, d^{-1}}$, at the considered latitude, both significantly larger than the solar values of $\alpha_{\odot} = 0.2$ and $\Delta\Omega_{\odot} = 0.066 \, \rm{rad\, d^{-1}}$, respectively.

Late-F-type stars are characterised by thin and shallow convection zones (CZs). As the CZ becomes shallower, the convective turnover time decreases compared to that in solar-type stars. This shorter turnover time enhances surface shear through the $\Lambda$-effect, potentially giving rise to solar-like differential rotation patterns if the Rossby number ($Ro = \frac{P_{rot}}{\tau}$) is sufficiently low. However, due to their shorter turnover times $\tau$, late-F stars must rotate considerably faster than the Sun to reach the same $Ro$, leading to the development of solar-type rotation and meridional flow patterns at much shorter rotation periods, accompanied by significantly stronger horizontal shear than observed on the solar surface \citep{kuker2007modelling}.

Both \cite{kuker2005differential} and \cite{balona2016differential} found that, in F-type stars, horizontal shear increases with increasing rotation rate. Our measured absolute horizontal shear of $\Delta\Omega = 0.49 \, \rm{rad\, d^{-1}}$ is consistent with these findings and agrees with \cite{ammler2012new}, who reported values between 0.1 and 1.0 $\rm{rad\, d^{-1}}$ for F8-type stars.

In contrast, \cite{balona2016differential} also showed that relative shear decreases steeply with increasing rotation rate, confirming the result of \cite{kuker2005differential} that relative shear peaks in solar-type stars. Our measured relative shear, $\alpha = 0.38$, aligns with \cite{reiners2003differential}, who observed that differential rotation with $\alpha > 0.1$ is common among slowly rotating F stars (vsini $\leq$ 50 km/s). Nonetheless, our $\alpha$ value exceeds the solar value and contradicts the lower predictions of \cite{kuker2005differential}. Notably, \cite{ammler2012new} also reported that $\alpha$ values in late-F stars tend to exceed theoretical predictions and occur at significantly shorter rotation periods, resulting in a stronger shear. They reported that the strongest amount of relative differential rotation ($54\%$) detected is found among F stars. These findings suggest that the shear observed in late-F stars consistently surpasses the levels predicted by current models.

\label{disc}
\section{Conclusions}
\label{conc}
In this study, we analysed the S-index data of HIP12653 and identified two magnetic cycle periods: $674.7$ d, corresponding to the inactive branch, and $5799.2$ d, associated with the active branch, which exhibits a decay phase longer than the rising phase. The residuals revealed a periodic variation of 4.8 d, attributed to a large number of measurements taken during the activity cycle's minimum. After excluding these measurements, we obtained a period of 7.7 days, which closely matches the rotation period reported in the literature.

To confirm that these two periods correspond to ARs at different latitudes, we performed a periodic signal analysis on TESS photometric data using an alternative period selection approach. 

We introduced a fitting model incorporating a defined number of harmonics to account for the potential presence of harmonics of the primary rotation period. Applying this model to a Monte Carlo simulation of stars with ARs with varying AR configurations, we achieved a $92.1\%$ success rate in estimating secondary periods within three times uncertainties.  It was also noted that smaller shears and periods close to the equatorial period were missed due to limitations in frequency resolution. 

Applying this method to HIP12653 TESS data across different sectors spanning 2018 to 2023, we retrieved the following periods: $4.8$ d, $5.7$ d, and $7.7$ d. These results confirm that the period of $4.8$ d detected in the S-index data corresponds to ARs at different latitude, closer to the equator. Assuming that the star exhibits a solar-like differential rotation, we estimate an inclination of approximately $34.0 \pm 1.8^{\circ}$ and a lower limit of the rotational shear of the activity belt of $\alpha = 0.38\pm 0.01$. It is worth mentioning that the stellar magnetic activity strongly depends on the Rossby number, which is related to the stellar rotation period. A more precise determination of the equatorial rotation period, in particular, and the amount of latitudinal differential rotation (shear) observed at the stellar surface are key parameters for dynamo models.

The large number of free parameters (inclination, ARs number, locations, and size) makes it challenging to model light curve variations. These parameters can introduce degeneracies, where multiple models with different AR configurations can produce the same light curve variations \citep{walkowicz2013information}. Therefore, a simpler approach to estimating differential rotation is to apply the GLS periodogram with a multi-harmonic model and cross-check the results using S-index data.

This study demonstrates that periodic searches conducted at various phases of the magnetic cycle can effectively reveal differential rotation. With sufficient measurements during the late phase of the magnetic cycle, the equatorial rotation period can be determined. 
Further observational campaigns can be very beneficial to provide greater clarity on these findings as more stellar magnetic cycles are uncovered.
\begin{acknowledgements}
This research was achieved using the POLLUX database ( \url{http://pollux.oreme.org} ) operated at LUPM  (Université Montpellier - CNRS, France) with the support of the PNPS and INSU. Montillet, J., Finsterle, W., 2024. PMO6v8 - Total Solar Irradiance Dataset from VIRGO/PMO6 , Version 1.0. Interdisciplinary Earth Data Alliance (IEDA). \url{https://doi.org/10.60520/IEDA/113532}. Accessed 2025-02-22. This paper includes data collected with the TESS mission, obtained from the MAST data archive at the Space Telescope Science Institute (STScI). Funding for the TESS mission is provided by the NASA Explorer Program. STScI is operated by the Association of Universities for Research in Astronomy, Inc., under NASA contract NAS 5–26555.
\end{acknowledgements}

\bibliographystyle{raa}
\bibliography{bibtex}

\begin{thebibliography}{65}
\providecommand\natexlab[1]{#1}
\providecommand\JournalTitle[1]{#1}

\bibitem[Alvarado-G{\'o}mez {et~al.}(2018)]{alvarado2018far}
Alvarado-G{\'o}mez, J.~D., Hussain, G.~A., Drake, J.~J., {et~al.} 2018, MNRAS, 473, 4326

\bibitem[Amazo-G{\'o}mez {et~al.}(2023)]{amazo2023far}
Amazo-G{\'o}mez, E.~M., Alvarado-G{\'o}mez, J., Poppenh{\"a}ger, K., {et~al.} 2023, MNRAS, 524, 5725

\bibitem[Ammler-von Eiff \& Reiners(2012)]{ammler2012new}
Ammler-von Eiff, M., \& Reiners, A. 2012, A\&A, 542, A116

\bibitem[Ara{\'u}jo {et~al.}(2025)]{araujo2025starspot}
Ara{\'u}jo, A., Lima, C., Menezes, F., \& Valio, A. 2025, The Astrophysical Journal Letters, 985, L28

\bibitem[Arenou {et~al.}(2023)]{arenou2023gaia}
Arenou, F., Babusiaux, C., Barstow, M.~A., {et~al.} 2023, A\&A, 674, A34

\bibitem[Baliunas {et~al.}(1995)]{baliunas1995chromospheric}
Baliunas, S., Donahue, R., Soon, W., {et~al.} 1995, Astrophysical Journal, Part 1 (ISSN 0004-637X), vol. 438, no. 1, p. 269-287, 438, 269

\bibitem[Balona \& Abedigamba(2016)]{balona2016differential}
Balona, L.~A., \& Abedigamba, O.~P. 2016, MNRAS, 461, 497

\bibitem[Barnes {et~al.}(2005)]{barnes2005dependence}
Barnes, J.~R., Cameron, A.~C., Donati, J.-F., {et~al.} 2005, MNRAS: Letters, 357, L1

\bibitem[B{\"o}hm-Vitense(2007)]{bohm2007chromospheric}
B{\"o}hm-Vitense, E. 2007, ApJ, 657, 486

\bibitem[Boisse {et~al.}(2012)]{boisse2012soap}
Boisse, I., Bonfils, X., \& Santos, N. 2012, A\&A, 545, A109

\bibitem[Boulkaboul {et~al.}(2022)]{boulkaboul2022analysis}
Boulkaboul, A., Damerdji, Y., Morel, T., {et~al.} 2022, MNRAS, 517, 1849

\bibitem[Bressan {et~al.}(2012)]{bressan2012parsec}
Bressan, A., Marigo, P., Girardi, L., {et~al.} 2012, Monthly Notices of the Royal Astronomical Society, 427, 127

\bibitem[Bruntt {et~al.}(2010)]{bruntt2010accurate}
Bruntt, H., Bedding, T.~R., Quirion, P.-O., {et~al.} 2010, MNRAS, 405, 1907

\bibitem[Carrington(1858)]{carrington1858distribution}
Carrington, R.~C. 1858, MNRAS, Vol. 19, p. 1-3, 19, 1

\bibitem[Charbonneau(2012)]{charbonneau2012solar}
Charbonneau, P. 2012, Solar and Stellar Dynamos: Saas-Fee Advanced Course 39 Swiss Society for Astrophysics and Astronomy, Vol.~39 (Springer)

\bibitem[Charbonneau(2020)]{charbonneau2020dynamo}
Charbonneau, P. 2020, Living Reviews in Solar Physics, 17, 1

\bibitem[Charbonneau \& Sokoloff(2023)]{charbonneau2023evolution}
Charbonneau, P., \& Sokoloff, D. 2023, Space Science Reviews, 219, 35

\bibitem[Collier~Cameron(2007)]{collier2007differential}
Collier~Cameron, A. 2007, Differential rotation on rapidly rotating stars

\bibitem[{de Bruijne} {et~al.}(2001)]{2001A&A...367..111D}
{de Bruijne}, J.~H.~J., {Hoogerwerf}, R., \& {de Zeeuw}, P.~T. 2001, \aap, 367, 111

\bibitem[De~Laverny {et~al.}(2012)]{de2012ambre}
De~Laverny, P., Recio-Blanco, A., Worley, C., \& Plez, B. 2012, A\&A, 544, A126

\bibitem[Duncan {et~al.}(1991)]{duncan1991ii}
Duncan, D.~K., Vaughan, A.~H., Wilson, O.~C., {et~al.} 1991, \apjs, 76, 383

\bibitem[Finsterle {et~al.}(2021)]{finsterle2021total}
Finsterle, W., Montillet, J.~P., Schmutz, W., {et~al.} 2021, Scientific reports, 11, 7835

\bibitem[Flores {et~al.}(2016)]{flores2016possible}
Flores, M.~G., Buccino, A.~P., Saffe, C.~E., \& Mauas, P.~J. 2016, MNRAS, 464, 4299

\bibitem[Foster(1996)]{foster1996wavelets}
Foster, G. 1996, Astronomical Journal v. 112, p. 1709-1729, 112, 1709

\bibitem[Fr{\"o}hlich {et~al.}(1995)]{frohlich1995virgo}
Fr{\"o}hlich, C., Romero, J., Roth, H., {et~al.} 1995, Solar Physics, 162, 101

\bibitem[Gustafsson {et~al.}(2008)]{gustafsson2008grid}
Gustafsson, B., Edvardsson, B., Eriksson, K., {et~al.} 2008, A\&A, 486, 951

\bibitem[Hall(2008)]{hall2008stellar}
Hall, J.~C. 2008, Living Reviews in Solar Physics, 5, 1

\bibitem[Hathaway(2007)]{hathaway2007solar}
Hathaway, D.~H. 2007, in American Astronomical Society Meeting Abstracts\# 210, Vol. 210, 99

\bibitem[Heck {et~al.}(1985)]{heck1985period}
Heck, A., Manfroid, J., \& Mersch, G. 1985, \aaps, 59, 63

\bibitem[I{\c{s}}{\i}k {et~al.}(2018)]{icsik2018forward}
I{\c{s}}{\i}k, E., Solanki, S.~K., Krivova, N.~A., \& Shapiro, A.~I. 2018, Astronomy \& Astrophysics, 620, A177

\bibitem[K{\"a}pyl{\"a} {et~al.}(2023)]{kapyla2023simulations}
K{\"a}pyl{\"a}, P.~J., Browning, M.~K., Brun, A.~S., Guerrero, G., \& Warnecke, J. 2023, Space Science Reviews, 219, 58

\bibitem[Katsova {et~al.}(2010)]{katsova2010differential}
Katsova, M., Livshits, M., Soon, W., Baliunas, S., \& Sokoloff, D. 2010, New Astronomy, 15, 274

\bibitem[{Kiehlmann} {et~al.}(2023)]{2023ascl.soft10003K}
{Kiehlmann}, S., {Max-Moerbeck}, W., \& {King}, O. 2023, {wwz: Weighted wavelet z-transform code}, Astrophysics Source Code Library, record ascl:2310.003

\bibitem[K{\"u}ker \& R{\"u}diger(2005)]{kuker2005differential}
K{\"u}ker, M., \& R{\"u}diger, G. 2005, Astronomische Nachrichten: Astronomical Notes, 326, 265

\bibitem[K{\"u}ker \& R{\"u}diger(2007)]{kuker2007modelling}
K{\"u}ker, M., \& R{\"u}diger, G. 2007, Astronomische Nachrichten: Astronomical Notes, 328, 1050

\bibitem[Lanza {et~al.}(2014)]{lanza2014measuring}
Lanza, A., Chagas, M.~D., \& De~Medeiros, J. 2014, A\&A, 564, A50

\bibitem[Lebreton {et~al.}(2001)]{lebreton2001helium}
Lebreton, Y., Fernandes, J., \& Lejeune, T. 2001, A\&A, 374, 540

\bibitem[Lenz \& Breger(2005)]{lenz2005period04}
Lenz, P., \& Breger, M. 2005, Communications in Asteroseismology, vol. 146, p. 53-136, 146, 53

\bibitem[{Lightkurve Collaboration} {et~al.}(2018)]{2018ascl.soft12013L}
{Lightkurve Collaboration}, {Cardoso}, J.~V.~d.~M., {Hedges}, C., {et~al.} 2018, {Lightkurve: Kepler and TESS time series analysis in Python}, Astrophysics Source Code Library, ascl:1812.013

\bibitem[Lovis {et~al.}(2011)]{lovis2011harps}
Lovis, C., Dumusque, X., Santos, N., {et~al.} 2011, arXiv preprint arXiv:1107.5325

\bibitem[McQuillan {et~al.}(2013)]{mcquillan2013measuring}
McQuillan, A., Aigrain, S., \& Mazeh, T. 2013, MNRAS, 432, 1203

\bibitem[Metcalfe {et~al.}(2010)]{metcalfe2010discovery}
Metcalfe, T., Basu, S., Henry, T., {et~al.} 2010, ApJL, 723, L213

\bibitem[Montes {et~al.}(2001)]{montes2001late}
Montes, D., L{\'o}pez-Santiago, J., G{\'a}lvez, M., {et~al.} 2001, MNRAS, 328, 45

\bibitem[Nielsen {et~al.}(2013)]{nielsen2013rotation}
Nielsen, M.~B., Gizon, L., Schunker, H., \& Karoff, C. 2013, Astronomy \& Astrophysics, 557, L10

\bibitem[Noyes {et~al.}(1984)]{noyes1984rotation}
Noyes, R., Hartmann, L., Baliunas, S., Duncan, D., \& Vaughan, A. 1984, \apj, 279, 763

\bibitem[Ol{\'a}h {et~al.}(2016)]{olah2016magnetic}
Ol{\'a}h, K., K{\H{o}}v{\'a}ri, Z., Petrovay, K., {et~al.} 2016, A\&A, 590, A133

\bibitem[Ol{\'a}h {et~al.}(2009)]{olah2009multiple}
Ol{\'a}h, K., Koll{\'a}th, Z., Granzer, T., {et~al.} 2009, A\&A, 501, 703

\bibitem[Radick {et~al.}(1995)]{radick199512}
Radick, R.~R., Lockwood, G., Skiff, B., \& Thompson, D. 1995, ApJ v. 452, p. 332, 452, 332

\bibitem[Reiners \& Schmitt(2003)]{reiners2003differential}
Reiners, A., \& Schmitt, J. 2003, Astronomy \& Astrophysics, 412, 813

\bibitem[Reiners \& Schmitt(2002)]{reiners2002feasibility}
Reiners, A., \& Schmitt, J.~H. 2002, A\&A, 384, 155

\bibitem[Reinhold \& Reiners(2013)]{reinhold2013fast}
Reinhold, T., \& Reiners, A. 2013, A\&A, 557, A11

\bibitem[Saar \& Osten(1997)]{saar1997rotation}
Saar, S., \& Osten, R. 1997, MNRAS, 284, 803

\bibitem[Sanz-Forcada {et~al.}(2013)]{sanz2013iotahorologi}
Sanz-Forcada, J., Stelzer, B., \& Metcalfe, T. 2013, A\&A, 553, L6

\bibitem[Sharma {et~al.}(2021)]{sharma2021differential}
Sharma, J., Kumar, B., Malik, A.~K., \& Vats, H.~O. 2021, Monthly Notices of the Royal Astronomical Society, 506, 4952

\bibitem[Solanki(2003)]{solanki2003sunspots}
Solanki, S.~K. 2003, The Astronomy and Astrophysics Review, 11, 153

\bibitem[Soubiran {et~al.}(2016)]{soubiran2016pastel}
Soubiran, C., Le~Campion, J.-F., Brouillet, N., \& Chemin, L. 2016, \aap, 591, A118

\bibitem[Soubiran {et~al.}(2018)]{soubiran2018gaia}
Soubiran, C., Jasniewicz, G., Chemin, L., {et~al.} 2018, \aap, 616, A7

\bibitem[Strassmeier(2009)]{strassmeier2009starspots}
Strassmeier, K.~G. 2009, The Astronomy and Astrophysics Review, 17, 251

\bibitem[Takeda(2020)]{takeda2020detection}
Takeda, Y. 2020, Publications of the Astronomical Society of Japan, 72, 10

\bibitem[Vaughan {et~al.}(1978)]{vaughan1978flux}
Vaughan, A.~H., Preston, G.~W., \& Wilson, O.~C. 1978, PASP, 90, 267

\bibitem[Vida {et~al.}(2014)]{vida2014looking}
Vida, K., Ol{\'a}h, K., \& Szab{\'o}, R. 2014, MNRAS, 441, 2744

\bibitem[Vogt \& Penrod(1983)]{vogt1983doppler}
Vogt, S.~S., \& Penrod, G.~D. 1983, Publications of the Astronomical Society of the Pacific, 95, 565

\bibitem[Walkowicz {et~al.}(2013)]{walkowicz2013information}
Walkowicz, L.~M., Basri, G., \& Valenti, J.~A. 2013, ApJS, 205, 17

\bibitem[Wauters {et~al.}(2016)]{wauters2016lyra}
Wauters, L., Dominique, M., \& Dammasch, I. 2016, Solar physics, 291, 2135

\bibitem[Zechmeister \& K{\"u}rster(2009)]{zechmeister2009generalised}
Zechmeister, M., \& K{\"u}rster, M. 2009, \aap, 496, 577

\end{thebibliography}

\label{lastpage}

\end{document}